\documentclass[
               ]{jetp} 

\usepackage{amsmath,amssymb,bm}
\usepackage{graphicx}
\usepackage[title]{appendix}

\newcommand{\angstrom}{\textup{\AA}}

\usepackage[square,numbers,sort&compress]{natbib}
\usepackage{color}
\begin{document}
\English
%
%
\title{Entropy signatures of topological phase transitions}

\author{Y.~M.}{Galperin}
\affiliation{Department of Physics, University of Oslo, P. O. Box 1048 Blindern, 0316 Oslo, Norway}
\affiliation{Ioffe Physical Technical Institute, 26 Polytekhnicheskaya, 
 St Petersburg 194021, Russian Federation}

\author{D.}{Grassano}
\affiliation{Dept. of Physics, and INFN, University of Rome Tor Vergata, Via della Ricerca Scientifica 1, I-00133 Rome, Italy}

\author{V.~P.} {Gusynin}
\affiliation{Bogolyubov Institute for Theoretical Physics, National Academy of Science of Ukraine, 
14-b Metrolohichna Street, Kiev 03680, Ukraine}

\author{A.~V.} {Kavokin}
\affiliation{CNR-SPIN, Viale del Politecnico 1, I-00133 Rome, Italy and Physics and Astronomy, University of Southampton, Higfield, Southampton, SO171BJ, UK}

\author{O.}{ Pulci}
\affiliation{Dept. of Physics, and INFN, University of Rome Tor Vergata, Via della Ricerca Scientifica 1, I-00133 Rome, Italy}

\author{S.~G.} {Sharapov}
\affiliation{Bogolyubov Institute for Theoretical Physics, National Academy of Science of Ukraine, 14-b Metrolohichna Street, Kiev 03680, Ukraine}

\author{V.O.}{ Shubnyi}
\affiliation{Department of Physics, Taras Shevchenko National University of Kiev,
6 Academician Glushkov ave., Kiev 03680, Ukraine}

\author{A.~A.} {Varlamov}
\affiliation{CNR-SPIN, Viale del Politecnico 1, I-00133 Rome, Italy}
\email{varlamov@ing.uniroma2.it}




\abstract{We review the behavior of the entropy per particle in various two-dimensional electronic systems. The entropy per particle is an important characteristic of any many body system that tells how the entropy of the ensemble of electrons changes if one adds one more electron. Recently, it has been demonstrated how the entropy per particle of a two-dimensional electron gas can be extracted from the recharging current dynamics in a planar capacitor geometry. These experiments pave the way to the systematic studies of entropy in various crystal systems including novel two-dimensional crystals such as gapped graphene, germanene and silicene. Theoretically, the entropy per particle is linked to the temperature derivative of the chemical potential of the electron gas by the Maxwell relation. Using this relation, we calculate the entropy per particle in the vicinity of topological transitions in various two-dimensional electronic systems. We show that the entropy experiences quantized steps at the points of Lifshitz transitions in a two-dimensional electronic gas with a parabolic energy spectrum. In contrast, in doubled-gapped Dirac materials, the entropy per particles demonstrates characteristic spikes once the chemical potential passes through the band edges. The transition from a topological to trivial insulator phase in germanene is manifested by the disappearance of a strong zero-energy resonance in the entropy per particle dependence on the chemical potential. We conclude that studies of the entropy per particle shed light on multiple otherwise hidden peculiarities of the electronic band structure of novel two-dimensional crystals.
}
\maketitle

\section*{Contents}
\JETPLTOC{section}{\LocalTOCitem{}{Introduction}{2}}
\JETPLTOC{section}{\LocalTOCitem{1}{General expressions}{4}}
\JETPLTOC{section}{\LocalTOCitem{2}{Quasi-two-dimensional electron gas: Quantization of entropy}{4}}
\JETPLTOC{section}{\LocalTOCitem{3}{ Lifshitz transitions in gapped Dirac materials: Entropy spikes}{9}}
\JETPLTOC{section}{\LocalTOCitem{4}{Entropy per particle in transition metal dichalcogenides}{14}}
\JETPLTOC{section}{\LocalTOCitem{5}{Entropy measurements as a tool for detection of topological transitions}{19}}
\JETPLTOC{section}{\LocalTOCitem{}{Conclusions}{25}}
\JETPLTOC{section}{\LocalTOCitem{}{Acknowledgements}{26}}
\JETPLTOC{section}{\LocalTOCitem{A}{Derivation of Equations~(\ref  {dndT}) and (\ref  {dndm})}{26}}
\JETPLTOC{section}{\LocalTOCitem{B}{Gapped Dirac materials: Details of calculations}{28}}
\JETPLTOC{section}{\LocalTOCitem{}{References}{29}}

\section*{Introduction} \label{introduction}

This review article is dedicated to Lev Petrovich Pitaevskii whose outstanding role in development of theoretical physics is well known. All the authors of this review had studied physics by the famous ``Course of Theoretical Physics'' written by Landau and Lifshitz with pronounced contribution of Lev Petrovich. This course, as well as his seminal journal articles, have shaped modern condensed matter physics.  We are happy to congratulate Lev Petrovich with his 85th birthday and wish him many years of health and fruitful work in theoretical physics.

Low-dimensional electronic structures are of a great interest because of their importance as the building blocks for quantum electronics.
Another reason is that size quantization of the electronic states in
low-dimensional systems leads to quantization of their thermodynamic and transport
properties. The most famous are the integer~\cite{vonKlitzing} and
fractional~\cite{Tsui82} quantum Hall effect in two-dimensional electron gas
(2DEG) and conductance quantization of quasi-one-dimensional channels~\cite{Wharam,VanWees}.

In the recent decades, a new class of atomically thin two-dimensional (2D) crystals has come to the focus of numerous experimental and theoretical studies. Starting from the pioneering works on the gapless 2D semimetal, graphene, the physics of 2D crystals advances at a high pace. A variety of gapped two-dimensional crystals has come into play, including, e.g., the transition metal dichalcogenides such as molybdenum or tungsten disulphites and diselenides. 

One of the ways to describe the anomalies of the properties of low-dimensional electron systems appearing in the result of size quantization is the formalism of Lifshitz topological transitions~\cite{Lifshitz1960JETP}. In three dimensional bulk crystals, Lifshitz transitions are sometimes referred to as $2\frac12$ order transitions. In contrast, in 2D crystals, this is no more universally valid, and particular cases need to be specifically analyzed. 
In accordance to the Ehrenfest terminology, the resonance of the electronic chemical potential with the energy level of size-quantization in a two-dimensional electron gas (2DEG) with a parabolic spectrum should be interpreted as the second order phase transition. The crossing points of the electronic chemical potential and the Landau level achieved by tuning the magnetic field perpendicular to the crystal plane may be identified as $1\frac12$ order transitions. Such transitions are accompanied by the steps or logarithmic singularities in the electronic density of states (DOS), openings of the specific channels in the electron scattering, giant resonances in thermoelectric power~\cite{BPV92}, de Haas-van Alphen and Shubnikov-de Haas oscillations ~\cite{Blanter1994PR}.
Manifestations of these phase transitions constitute an important part of the modern physics of two-dimensional electronic systems.

In recent years, a new class of topological materials has been theoretically predicted and experimentally studied (see reviews~\cite{Hasan2010RMP,Qi2011RMP}).
Topological insulators are characterized by bulk band gaps and gapless edges or surface states, that are protected by the time-reversal symmetry and characterized by a $Z_2$ topological order parameter.
Novel group-IV graphene-like two-dimensional crystals such as silicene, germanene and stanene are examples of the two-dimensional topological insulators proposed in \cite{Kane.Mele.2005,Bernevig2006PRL}.
They attract an enhanced attention nowadays because of their high potential for applications in nanoelectronic devices of a new generation.

In order to study the peculiar electronic properties of both classical semiconductor quantum wells and novel two-dimensional crystals, one needs an experimental technique that would be sensitive to the peculiarities of the electronic band structure. Traditional methods, such as optical transmission,  sometimes fail to detect the variations of the electronic DOS in the far-infrared and terahertz spectral range~\cite{bechstedt2012infrared,Stille2012PRB}, while electronic conductivity measurements, even at ultra-low temperatures, provide only an indirect information on the band and spin structure of the studied material. 

Recently, a promising tool for the band structure studies with use of the electronic transport measurements has been proposed and successfully tested on a two-dimensional electron gas with a parabolic dispersion~\cite{Pudalov}.
This method, based on the measurements of recharging currents in a planar capacitor geometry, gives access to the entropy per particle $s \equiv (\partial S/ \partial n)_T$ ($S$ is the entropy per unit volume and $n$ is the electron density) at temperature $T$, which we shall express in energy units.
This characterization technique is based on the Maxwell relation that links the temperature derivative of the chemical potential in the system, $\mu$, to the entropy per particle $s$:
\begin{equation} \label{MR}
(\partial S/ \partial n)_T =
- (\partial \mu /\partial T)_n.
\end{equation}
The modulation of the sample temperature changes the chemical potential and, hence, causes recharging of the gated structure, where the 2D electron gas and the gate act as two plates of a capacitor.
Therefore, $(\partial \mu /\partial T)_n$ may be directly obtained in this experiment from the measured recharging current.

The entropy per particle is an important characteristic \textit{per se} of any many-body system.
It also governs the thermoelectric and thermomagnetic properties of the system entering explicitly the expressions for the Seebeck and Nernst-Ettingshausen coefficients~\cite{Varlamov2013EPL,goupil2011thermodynamics}.
In this Review, we address theoretically this major thermodynamic quantity -- entropy per particle -- of
quasi-two-dimensional electron gases in various solid state systems. We specifically focus on the behavior of entropy in the vicinity of topological transitions of different types.

The paper is organized as follows.
We start with general expressions for the entropy per particle, Section~\ref{Gen}.
In Section~\ref{2D}, we provide the general equations that link the entropy per particle with DOS, chemical potential and temperature.
We consider the specific case of a two dimensional electron gas characterized by parabolic energy subbands, as it is the case e.g. in semiconductor quantum wells. 
We show that the
quantization of the energy spectrum of quasi-2DEG with a parabolic dispersion into subbands leads to a
very specific quantization of the entropy: $s$ exhibits sharp maxima as the
chemical potential $\mu $ passes through the bottoms of size quantization
subbands ($E_{j}$). The value of the entropy in the $N$-th maximum depends 
\textit{only} on the size-quantization quantum number corresponding to this maximum, $N$: 
\begin{equation}  \label{s}
s\vert_{\mu = E_n} \equiv \left(\frac{\partial S}{\partial n}\right)_{T,\,
\mu = E_n} = \frac{\ln 2}{N - 1/2}\, .
\end{equation}
In the absence of scattering this result is independent of the shape of the
transverse potential that confines 2DEG and of the material parameters
including the electron effective mass and dielectric constant.
We reveal the quantization of entropy per electron at resonances of the chemical
potential and electron quantization levels and discuss the accuracy of the obtained expression for the quantized entropy in the presence of disorder and electron-electron interactions.

In Section~\ref{Dirac_materials},  we calculate the entropy per particle in the vicinity of topological transitions in various two-dimensional electronic systems. 
In contrast to the case of a two-dimensional electronic gas with a parabolic energy spectrum considered in Section~\ref{2D}, in double-gapped Dirac materials, the entropy per particle demonstrates characteristic spikes once the chemical potential passes through the band edges. We consider specific cases of the bilayer graphene and silicene.
We show that studies of the entropy per particle shed light on multiple otherwise hidden peculiarities of the electronic band structure of novel two-dimensional crystals.

In Section~\ref{dichalcogenides}, we focus on monoatomically thin layers of transition metal dichalcogendes. This class of structures is characterized by significant energy gaps that can be tuned by application of external fields and strain. This results in a peculiar band-structure that finds its unique signature in the entropy-per-particle dependence on the chemical potential. Particularly, we study the strain effect on the entropy per particle dependence on the chemical potential, and show that it may be very prominent even at elevated temperatures.

In Section~\ref{germanene}, we show that a specific resonant feature in the entropy per electron dependence on the chemical potential may be considered as a fingerprint of the transition between topological and trivial insulator phases in germanene.
The entropy per electron in a honeycomb two-dimensional crystal of germanene subjected to the external electric field is obtained from the first principle calculation of the density of electronic states and the Maxwell relation. We demonstrate that, in agreement to the recent prediction of the analytical model, strong spikes in the entropy per particle dependence on the chemical potential appear at low temperatures.
They are observed at the values of the applied bias both below and above the critical value that corresponds to the transition between the topological insulator and trivial insulator phases, while the giant resonant feature in the vicinity of zero chemical potential is strongly suppressed at the topological transition point, in the low temperature limit. In a wide energy range, the van Hove singularities in the electronic density of states manifest themselves as zeros in the entropy per particle dependence on the chemical potential.

In Conclusions, we outline the variety of problems linked to the entropy measurements and their interpretation that still remain to be solved. Some details of the calculations are elucidated in Appendices.

\section{General expressions} \label{Gen}

Let us recall that the entropy per particle is rather sensitive to the electron-hole asymmetry (see numerator of Eq.~(\ref{entropy-part-b}))
and plays an important role for the detection of the Lifshitz topological transition \cite{Blanter1994PR}. Whereas in the density of states (DOS) or conductivity the Lifshitz transitions manifest themselves as weak cusps,
in the thermoelectric power, directly related to $s(\mu,T)$, giant singularities occur. Such a singularity can be considered as a smoking gun of the topological Lifshitz transition.
Intuitively, one can expect that the sensitivity of $s(\mu,T)$  to topological transformations of the Fermi surface would give an opportunity to trace out various types of topological transitions. We will demonstrate that $s(\mu,T)$ indeed offers the opportunity to identify the transition between the topological and the trivial insulator, while the entropy features characterizing this transition are quite different from those characteristic for the Lifshitz transition.

The termodynamic coefficients of an electron system can be conveniently expressed through the electron density of states given by
\begin{equation}
	\label{eqn:DOS}
    D(\varepsilon) = \sum_{j=1}^M \int_{BZ} \frac{d{\bf k}}{(2\pi)^2} \delta \left( \varepsilon - \varepsilon_{j,{\bf k}} \right) ,
\end{equation}
with $j$ running over all the occupied bands up to the $M$th one.
The presence of disorder leads to smearing of the $\delta$-function. Therefore, in general it is a smeared $\delta$-function, $\delta_\gamma \left( \varepsilon - \varepsilon_{j,{\bf k}} \right)$, that enters Eq.~(\ref{eqn:DOS}).

For a Fermi system, the relationship between the electron concentration $n$, the chemical potential $%
\mu $ and temperature $T$ can be found integrating the product of DOS given by Eq.~(\ref{eqn:DOS}) and the Fermi-Dirac distribution function over the electron energy:
\begin{equation}
n\left( \mu ,T\right) =\int\limits_{0}^{+\infty }\frac{ D(\varepsilon)}{\exp \left( 
\frac{\varepsilon-\mu }{T}\right) +1}\,d\varepsilon.  \label{ngen-a}
\end{equation}
In its turn, the entropy per electron, $s$,   can be related to the function  $D (\varepsilon )$  using the Maxwell relation~(\ref{MR}):
\begin{equation}
s=\left( \frac{\partial S}{\partial n}\right) _{T}=-\left( \frac{\partial
\mu }{\partial T}\right) _{n}=\left( \frac{\partial n}{\partial T}\right)
_{\mu }\left( \frac{\partial n}{\partial \mu }\right) _{T}^{-1}.
\label{fullderiv}
\end{equation}%
Differentiation of Eq.~(\ref{ngen-a}) over temperature and chemical potential, respectively, results in the  general relation \cite{Varlamov2016PRB,Tsaran2017}:
\begin{equation} \label{entropy-part-b}
s (\mu,T)
=\frac1T\frac{\int _{-\infty}^{\infty }d \varepsilon\,  D (\varepsilon ) (\varepsilon - \mu) \cosh^{-2}
\left( \frac{\varepsilon - \mu}{2T} \right) }{\int _{-\infty}^{\infty }d \varepsilon D (\varepsilon ) \,  \cosh^{-2} \left( \frac{\varepsilon - \mu}{2T} \right) },
\end{equation}which we will exploit below for different systems  of the increasing DOS energy dependence complexity.

\section{Quasi-two-dimensional electron gas: Quantization of entropy} \label{2D}
\paragraph{Entropy per particle in quasi-2DEG. --}

In the absence of scattering, the density of electronic state in a
non-interacting 2DEG has a staircase-like shape~\cite{Ando}, 
\begin{equation}
D(\varepsilon )=\frac{m^{\ast }}{\pi \hbar ^{2}}\sum\limits_{j=1}^{\infty }\theta
\left(\varepsilon -E_{j}\right) ,  \label{nuclean}
\end{equation}%
with $m^{\ast }$ being the electron effective mass and $\theta \left(
x\right) $ being the Heaviside theta-function. Elastic scattering of
electrons against defects and impurities that are necessarily present in realistic
systems, leads to the smearing of the steps of the density of states. A
simple way to account for this smearing is to introduce a finite life-time, $\hbar/\gamma$, 
of an electron. That results in the replacement of the Dirac delta-function by the Lorentzian in the derivative of the density of states: 
\begin{equation} \label{tilde_delta}
\theta^{\prime }(\varepsilon)=\delta (\varepsilon)\rightarrow \delta_\gamma(\varepsilon)\equiv \frac{\gamma}{\pi (\varepsilon^{2}+\gamma^2)}. 
\end{equation}
Integration of the latter expression leads to the
replacement 
$
\theta (\varepsilon)\rightarrow \theta_\gamma(\varepsilon),
$
where 
\begin{equation}
\theta_\gamma\left( \varepsilon\right) =\frac{1}{2}+\frac{1}{\pi }\arctan \left( 
\frac{\varepsilon}{\gamma }\right) .  \label{theta}
\end{equation}%
We focus on the physically important limit of $T\gg \gamma $, that corresponds to a
relatively clean sample. At the same time, the temperature is supposed to be
not too high, $T\ll \Delta _{Nj}=|E_{N}-E_{j}|,\forall j\neq N$. In
addition, we assume that the transport is adiabatic~\cite{Beenakker}, i.e.,
there is no elastic inter-band transitions due to backscattering.

The relationship between the electron concentration $n$, the chemical potential $%
\mu $ and temperature $T$ can be found integrating the product of DOS given by Eq. (\ref{nuclean}) and the Fermi-Dirac distribution function over energy. The integration may be performed accounting for the
renormalization (\ref{theta}): 
\begin{equation}
n\left( \mu ,T\right) =\frac{m^{\ast }}{\pi \hbar ^{2}}\sum\limits_{i=1}^{%
\infty }\int\limits_{0}^{+\infty }\frac{\theta_\gamma(\varepsilon-E_{i})}{\exp \left( 
\frac{\varepsilon-\mu }{T}\right) +1}\,d\varepsilon.  \label{ngen}
\end{equation}%
Calculating the partial derivatives of the electron concentration over temperature and chemical potential one can express them in the form of sums over the subbands, which can be cast into the form (see Appendix A): 
\begin{eqnarray}
\left(\frac{\partial n}{\partial T}\right)_\mu
&=&\frac{2m*}{\pi^2\hbar^2}\sum_j{\rm Re}\left[\frac{\gamma+i(\mu-E_j)}{2T}\Psi\left(\frac{1}{2}+\frac{\gamma+i(\mu-E_j)}{2\pi T}\right)
-\frac{\gamma}{2T}\right.\nonumber\\
&&\left. \quad - \pi  \ln \left\{\Gamma\left(\frac{1}{2}+\frac{\gamma+i(\mu-E_j)}{2\pi T}\right)\right\}+\frac{\pi }{2} \ln(2\pi)\right],
\label{dndT}
\\
\left(\frac{\partial n}{\partial\mu}\right)_T &=&\frac{m*}{2\pi\hbar^2}\sum_i
\left\{1+\frac{2}{\pi}{\rm Im}\left[\Psi\left(\frac{1}{2}+\frac{\gamma+i (\mu-E_i)}{2\pi T}\right)\right]\right\}. \label{dndm} 
\end{eqnarray}
Here $\Psi (z)\equiv d \ln [\Gamma (z)]/dz$ is the digamma function that is a logarithmic derivative of the Euler $\Gamma$-function $\Gamma (z)$.
We have taken into account that \ $\mu \gg T$ and
extended the lower limit of integration down to $-\infty $.
The general expression Eq. (\ref{entropy-part-b})  for the entropy per particle in the quasi-2DEG can be calculated form the above expressions using Eq.~(\ref{fullderiv}).

The proposed replacement of the Dirac delta-function by a Lorentzian is a model assumption. Yet, already basing on dimensionality analysis one can see that any other way of delta-function smearing leads to the result qualitatively similar to what Eqs.~(\ref{dndT}) - (\ref{dndm})  describe. For example, replacing the delta-function by the Gaussian function one can easily find that Eq. (9) converts to the probability integral $\mbox{erf}(\epsilon/\gamma)$. This would not change qualitatively the picture in comparison with the Lorentzian smearing considered here, but it would prevent us from obtaining the explicit analytical results given by Eqs.~(\ref{dndT}) - (\ref{dndm}).

\paragraph{Quantization of the entropy per particle in a non-interacting 2DEG. --}

We start the analysis with the case of a clean material, where one can neglect the smearing of electron states by tending $\gamma \to 0$.
In this case the principal contribution to
the derivative (\ref{dndT}) is provided by the level closest to the chemical
potential: 
\begin{equation*}
\left( \frac{\partial n}{\partial T}\right) _{\mu \rightarrow E_{N}}=\frac{%
m^{\ast }}{\pi \hbar ^{2}}\left[ \ln \left( 2\cosh \frac{\delta _{N}}{2T}%
\right) -\frac{\delta _{N}}{2T}\tanh \frac{\delta _{N}}{2T}\right] .
\end{equation*}%
Here $\delta _{N}=\mu -E_{N}$, $N$ stands for the level closest to the chemical potential $\mu$ (we assume $|\delta _{N}|\ll \Delta _{N,N \pm 1}$ where $\Delta_{ij} \equiv |E_i - E_j|$).
The contributions of other levels are exponentially weak, they turn to be
of the order of  $\exp (-\Delta _{N,N\pm 1}/T)$. In 
Eq.~(\ref{dndm}), the
lowest $N-1$ levels provide the same constant, independent of the chemical
potential and temperature contributions, while the shape of the line is
governed by the N-th level. We obtain: 
\begin{equation}
\left( \frac{\partial n}{\partial \mu }\right) _{\mu \rightarrow E_{N}}=%
\frac{m^{\ast }}{\pi \hbar ^{2}}\left( N-\frac{1}{2}\right) +\frac{m^{\ast }%
}{2\pi \hbar ^{2}}\tanh \frac{\delta _{N}}{2T}.  \notag
\end{equation}%
The contributions of the higher energy quantization subbands ($j>N$) are exponentially small.

\begin{figure}[th]
\centering
\includegraphics[width=8cm]{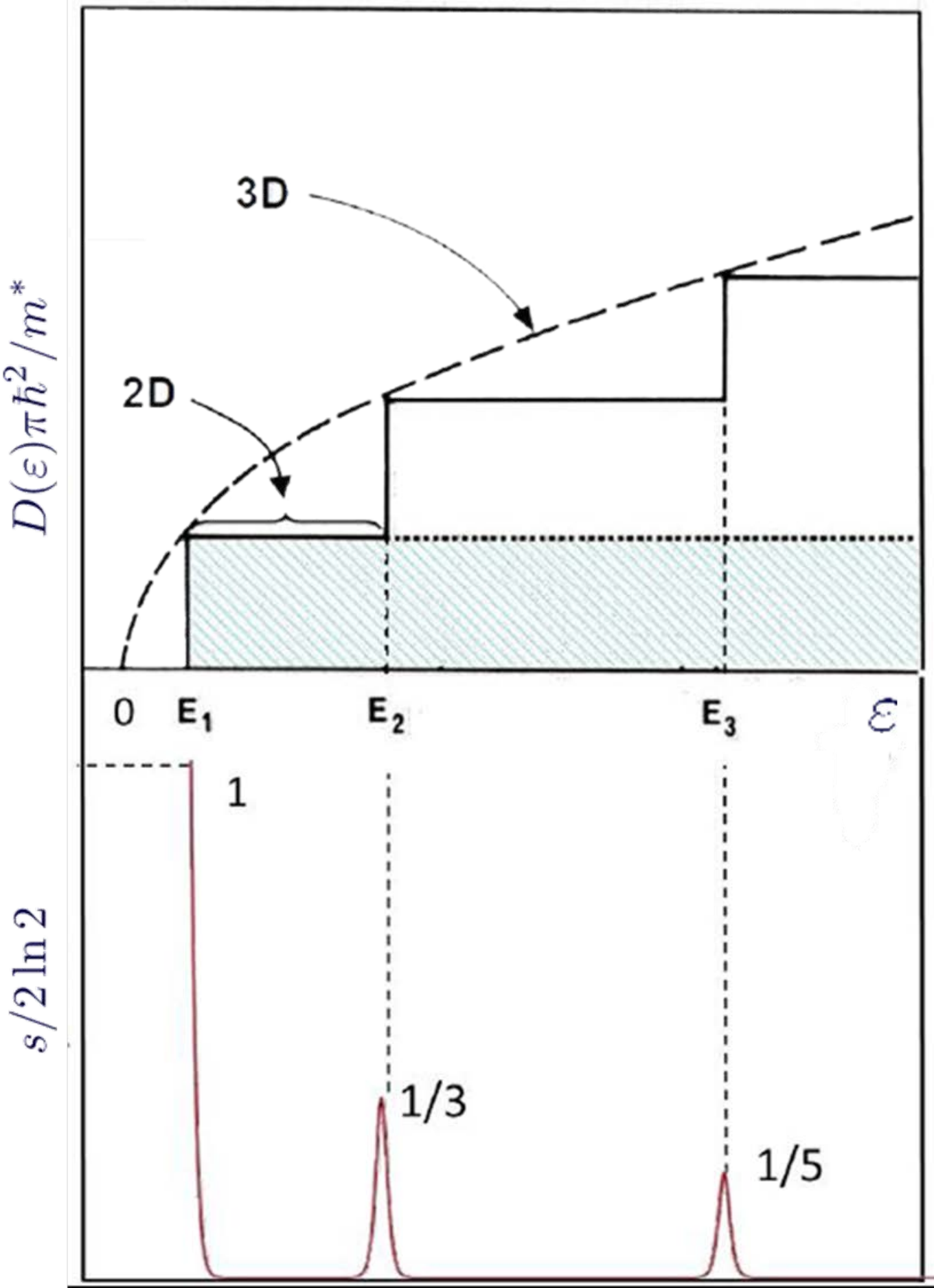}
\caption{Schematic representation of the dependencies of the electronic
density of states (upper panel) and the entropy per electron (lower panel) as
functions of the chemical potential.}
\label{fig1}
\end{figure}

Finally, the expression for the entropy per electron Eq. (\ref{fullderiv}), valid 
for any spectrum of size quantization  $E_{j},$\ takes the form: 
\begin{eqnarray}  \label{fin}
s_{\mu \rightarrow E_{N}} &=&\frac{\ln \left( 2\cosh \frac{\delta _{N}}{2T}%
\right) -\frac{\delta _{N}}{2T}\tanh \frac{\delta _{N}}{2T}}{\left(
N-1/2\right) +\frac{1}{2}\tanh \frac{\delta _{N}}{2T}}  \label{final} \\
&=&\left\{ 
\begin{array}{cc}
\frac{|\delta _{N}|}{T}\frac{\exp \left( -\frac{|\delta _{N}|}{T}\right) }{%
N-1+\exp \left( -\frac{|\delta _{N}|}{T}\right) }, & \delta _{N}\ll -T, \\ 
\frac{\ln 2}{N-1/2}, & 0<\delta _{N}\ll T, \\ 
\frac{\delta _{N}}{TN}\exp (-\delta _{N}/T), & \delta _{N}\gg T.%
\end{array}%
\right.   \label{finas}
\end{eqnarray}%
This expression predicts the existence of quantized peaks of the partial
entropy $s$ at $\mu =E_{N}$, their magnitudes being dependent only on the
quantization subband quantum number. The dependence of $s$ on the chemical potential is
schematically shown in Fig.~\ref{fig1}, lower panel. The quantized peaks of
the entropy per electron correspond to the steps of the density of states shown
in the upper panel of the same figure. 
The shape of the peaks in Eq. (\ref{final}) is asymmetric, that corresponds
to the step-like jumps in the dependence of the density of electronic states on the chemical potential of the 2DEG.

\paragraph{Effect of scattering. --}

Equations (\ref{dndT}) and (\ref{dndm}) allow one to estimate the effect of
scattering on the heights of the peaks. A straightforward analysis shows
that the contribution of the lower subbands is of the order of $%
\sum\limits_{j=1,j\neq N}^{\infty }\mathcal{O}\left[ \left( \Delta _{Nj}\tau
/\hbar \right) ^{-k}\right] $. Here $k=1$ for $(\partial n/\partial \mu )_{T}
$ and $k=2$ for $(\partial n/\partial T)_{\mu }$. These sums are cut off at $%
j_{\max }\sim \hbar /T\tau .$ For the equidistant spectrum (parabolic
potential), or $E_{N}\sim N^{2/3}$ (eigenvalues of the Airy functions, in
the case of the linear potential) they give small contributions of the order 
$\left( \Delta \tau /\hbar \right) ^{-k}\left( T/\Delta \right) ^{k-1}$($%
\Delta $ is the characteristic scale of inter-level distances).

As it follows from Eqs.~(\ref{dndm}) and (\ref{lim01}), the contribution of the $N$-th subband to $(\partial n/\partial \mu)_T$ can
be calculated exactly  leading to the replacement  
\begin{equation*}
\tanh \frac{\delta _{N}}{2T} \to 
\mathop{\mathrm{Re}}\left[\tanh \left(\frac{
\delta _{N}}{2T} -i\frac{\gamma}{2T}\right ) \right]
\end{equation*}
in Eq.~(\ref{final}), 
i.e., leading to the appearance of the corrections of the order $%
\mathcal{O}\left[ \left( \hbar /T\tau \right)^{2}\right] $. Yet, the
dominant effect of impurities is through the factor $(\partial n/\partial T)_\mu$. The
asymptotic analysis of Eq. (\ref{dndT}) shows that the magnitude of the peak
in $s$ is suppressed by the elastic scattering of electrons as%
\begin{equation*}
s_{\mu =E_{N}}=\frac{\ln 2-\left(\gamma / \pi T\right)}{N-1/2}.
\end{equation*}%
This simple relation allows to extracte the degree of disorder in a 2DEG from the entropy per particle dependence on the chemical potential.

\paragraph{The role of electron-electron interaction. --}

Now let us briefly discuss the role of electron-electron (e-e) interactions,
which are neglected in the above formalism. Electron-electron interactions become
noticeable for the electronic states sufficiently close to the subbands'
bottoms. In particular, they can significantly change the compressibility of the
electron gas, see, e.g.,~\cite{Eisenstein94}. 
Characterizing the strength of e-e interaction by a parameter $%
r_{s}$~\cite{Mahan} we conclude that $r_{s}\approx 1$ for the upper filled
subband at 
\begin{equation*}
\sqrt{\frac{2\delta _{N}}{m^{\ast }}}\approx \frac{e^{2}}{\kappa \hbar }%
\rightarrow \delta _{N}\approx \frac{m^{\ast }}{2}\left( \frac{e^{2}}{\kappa
\hbar }\right) ^{2},
\end{equation*}%
with $\kappa $ being a dielectric constant. \ Putting $m^{\ast }=0.1m_{0}$ and $\kappa =10$ we
get $\delta _{N}\gtrsim 2\times 10^{-14}$~erg. If $\delta _{N}$ is less than
this value one can expect a Fermi-liquid-induced renormalization of the electronic
spectrum, that, in particular, affects the effective mass $m^{\ast }$. Fortunately, $%
m^{\ast }$ doesn't enter the expressions for the magnitudes of the peaks of $s$. However,
an additional correction proportional to $(\partial m^{\ast }/\partial n)$ can
appear in the expression \eqref{dndm} for the thermodynamic DOS. These
corrections for $N=1$ are discussed in Ref.~\cite{Pudalov} and
references therein. Another possibility of evidencing the interaction
effects is establishing a special regime of a correlated 2D charged plasma~%
\cite{Novikov} also explored in Ref.~\cite{Pudalov}. Here we do not
consider this particular case. In general, the comparison of experimental
results with the universal expression ~\eqref{fin} obtained here would allow one to
judge on the role of electron-electron correlations in the system under study.

\paragraph{Physical interpretation of the entropy peaks. --}

The dependence of the entropy per electron $s$ on the chemical potential can be
interpreted in the following way. At low temperatures, the main contribution
to the entropy is provided by the electrons having energies in the vicinity
of the Fermi level, the width of the `active' energy range being $\sim T$. If the
electron DOS is constant within the active range, by adding an electron one
does not change the entropy. Hence, the entropy is independent of the
chemical potential, $(\partial S/\partial n)_{T}\rightarrow 0$. However, if
the bottom of one of the subbands falls into the active energy range, the number of
`active' states becomes strongly dependent on the chemical potential. In
this case, by adding an electron to the system, one changes the number of
'active' states in the vicinity of the Fermi surface. Consequently, the
entropy per electron strongly increases. The peaks of the entropy per electron
correspond to the resonances of the chemical potential and electron size
quantization levels. The further increase of the chemical potential brings
the system to the region of the constant density of states, where the
entropy per electron vanishes again.

The intersections by the chemical potential of the levels of electron size
quantization, $\delta _{N}=0$, can be considered as the points of
Lifshitz phase transitions in a 2DEG, where the Fermi surface acquires a
new component of topological connectivity. Corresponding anomalies in the
thermodynamic and transport characteristics, in particular, of the thermoelectric
coefficient related to the peculiarities of the energy dependence of the
electron momentum relaxation time have been studied experimentally and
theoretically in Refs.~\cite{Z1,Z2} and \cite{BPV92}, respectively. Here we
present an analytical theory of the entropy anomalies of a purely thermodynamic origin.

In the asymptotic expression \eqref{finas} for strongly negative $\delta_N $
(but $|\delta_N| \gg T$) , the item $\exp(-|\delta_N|/T)$ in the denominator
can be neglected for all $N>1$. However, it becomes important for $N=1$.
Indeed, at $\mu < E_1$ the entropy per electron increases as $|\mu -E_1|/T$ with
decreasing $\mu$. This is a manifestation of the crossover from the Fermi
distribution to the Boltzmann one where the chemical potential falls into the energy 
gap in the spectrum. The region $\mu < E_1$ is not shown in Fig.~\ref{fig1}
in order to keep the peaks for $N=2,3$ visible. We shall discuss the behaviour of the entropy per particle in the vicinity of zero chemical potential in Section 5.

At $T\rightarrow 0$ (yet $T\gtrsim \gamma $) the peaks of $s$ are
located at $\mu \rightarrow E_{N}$, $N>1$, the maximum values being $s_{\max
}(N)=\ln 2/(N-1/2)$. At finite $T$ the peaks acquire finite widths of the
order of $T$ and shift toward negative values of $\mu -E_{N}$. 
\begin{figure}[th]
\centering
\includegraphics[width=.7\columnwidth]{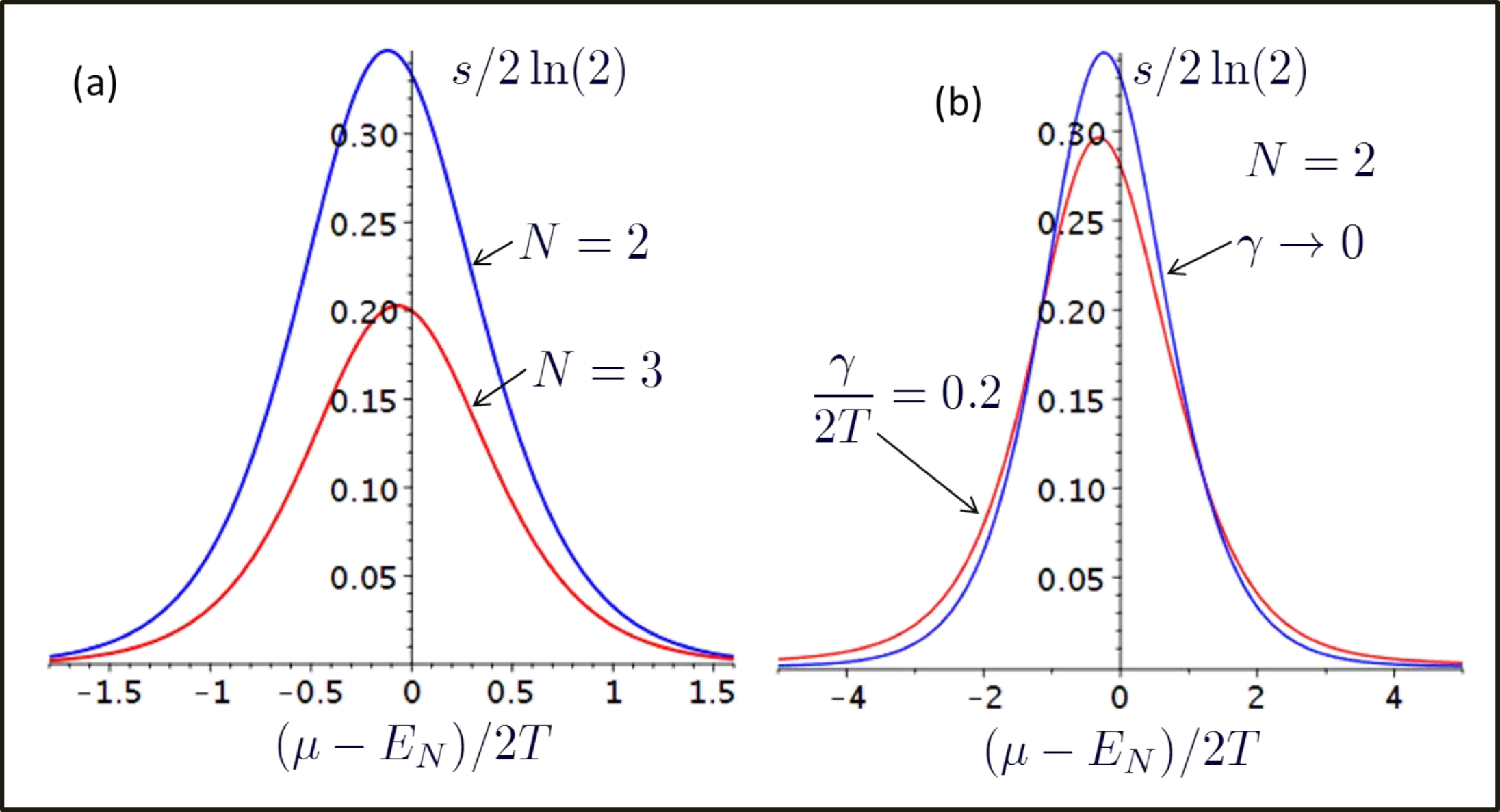}

\caption{(Color online) Dependence of the entropy per particle, $s$, on ($\mu - E_N)/2T$ for (a) $N=2,3$; $\gamma \to 0$; (b)  $N=2$; $\gamma/2T = 0, 0.2$.
\label{fig2}}
\end{figure}
The peaks of $s$ for $N=2$ and 3 and $\gamma \to 0$ are shown in Fig.~\ref{fig2}~(a), their
characteristics are given in Table~\ref{tab}. The reason of the peaks'
asymmetry is the difference in partial densities of states above and below
the chemical potential. The relative difference between the DOSs decreases
with increase of $N$, therefore the peaks become more symmetric at the crossections of the chemical potential with higher size-quantization levels. 
\begin{table}[h]
\centering
\begin{tabular}{|c|c|c|c|}
\hline
N & $\delta_{\max}/2T$ & $s_{\max}/2\ln 2$ & $s\vert_{\mu = E_N}/2\ln 2$ \\ 
\hline
2 & -0.24 & 0.347 & 1/3 \\ \hline
3 & -0.14 & 0.203 & 1/5 \\ \hline
4 & -0.01 & 0.144 & 1/7 \\ \hline
\end{tabular}%
\caption{Peaks in the entropy per electron. }
\label{tab}
\end{table}
The role of disorder is demonstrated for $N=2$ in Fig.~\ref{fig2}(b).

Interestingly, our result for $(\partial \mu/\partial T)_n \equiv -s$ at $%
\mu = E_1$ and $\gamma \to 0$ coincides with the expression for the same
derivative obtained in Refs.~\cite{Bob71,Abrikosov} for a two-dimensional
superconductor. The quantized dip in this derivative is associated with the
step in the density of electronic states which changes from zero inside the
superconducting gap to $m^{\ast }/\pi \hbar ^{2}$ above the gap. A
remarkable fact is that the value of the effective mass $m^{\ast }$ does not
enter the result.

Note that, in addition to the recharging current measurements discussed above~\cite{Pudalov}, the variation of the chemical potential as a function of
temperature can be measured by resonant optical transmission spectroscopy of
the fundamental absorption edge in modulation doped semiconductor quantum
wells, see, e.g.,~\cite{Dutton1958}.

\section{
Lifshitz transitions in gapped Dirac materials: Entropy spikes} \label{Dirac_materials}

In this Section, we analyze theoretically the behavior of
the entropy per particle as a function of the chemical potential in a gapped graphene deposited on a substrate and in other low-buckled Dirac materials, e.g., silicene and germanene.
We show that  the entropy per electron in these systems acquires quantized universal values
at low temperatures if the  chemical potential passes through  the edges of corresponding gaps.
This feature will be shown to be a universal property of electronic systems characterized by a step-like behavior of the density of states. In this sense the behavior is similar to that we analyzed for the case of a quasi-2D electron gas in the previous Section.
If the chemical potential is resonant to the Dirac point, we find the discontinuity in $s$ at zero temperature.
At low but finite temperatures this discontinuity transforms into the combination of a very sharp dip at the negative chemical potential followed by a sharp peak at the positive chemical potential.
These predictions offer a new tool for the characterization of novel crystalline structures.
In particular, the very characteristic spikes of entropy that must be relatively easy to observe
are indicative of the energy gaps appearing, in particular, due to the the spin-orbit interaction. We believe that the measurements of the entropy per particle (e.g., following the technique of Ref. \cite{Pudalov}) may reveal hidden peculiarities of the band structure of new 2D materials.

\paragraph{Discontinuities of DOS and quantization of entropy. --}

In order to describe gapped Dirac materials we assume that the DOS is a symmetric function,
$D (\varepsilon) =D (-\varepsilon)$, and it has $2N$ discontinuities at the points $E = \pm\Delta_i$.
Thus it can be written in the form
\begin{equation}
\label{DOS-general}
D\left( \varepsilon\right) =f(\varepsilon)\sum_{i=1}^M \theta \left(\varepsilon^{2}-\Delta_i^{2}\right).
\end{equation}
The function $f(\varepsilon)$ is assumed to be a continuous even function of energy $E$, and it may
account for the renormalizations due to electron-electron interactions in the system.

The case of $M=1$ corresponds to a gapped graphene with the dispersion dependence $\varepsilon (k) = \pm \sqrt{\hbar ^{2} v_{F}^{2} k^{2} + \Delta ^{2}}$
and $f(\varepsilon) = 2 |\varepsilon|/(\pi  \hbar ^{2} v_{F}^{2})$, where we have taken into consideration both the valley and spin degeneracy.
Here $ \Delta $ is the energy gap, $v_{F}$ is the Fermi velocity, $\mathbf{k}$ is the wave vector. The global sublattice asymmetry gap $2 \Delta \sim 350$~K can be introduced in graphene~\cite{Hunt2013Science,Woods2014NatPhys,Chen2014NatCom,Gorbachev2014Science} if it is placed on top of a hexagonal boron nitride (G/hBN) and the crystallographic axes of graphene and hBN are aligned.

The specific case of $M=2$ corresponds to
silicene~\cite{Kara2012SSR}, germanene~\cite{Acun2015JPCM} and other low-buckled Dirac materials~\cite{Liu2011PRL,Liu2011PRB}.
The dispersion dependence in these materials writes 
\begin{equation} \label{silicene-spectrum}
\varepsilon_{\eta \sigma} (k) = \pm \sqrt{\hbar ^{2} v_{F}^{2} k^{2} + \Delta_{\eta \sigma} ^{2}},
\end{equation}
where $\eta$ and $\sigma$ are the valley and spin indices, respectively. Here the valley- and spin-dependent gap,
$\Delta _{\eta \sigma }=\Delta _{z}-\eta \sigma \Delta _{\text{SO}}$, where $\Delta _{\text{SO}}$ is the material-dependent spin-orbit gap induced by a strong intrinsic spin-orbit interaction.
It may have a relatively large value, e.g., $\Delta _{\text{SO}} \approx {4.2}$~{meV} in silicene and
$\Delta _{\text{SO}} \approx {11.8}$~{meV} in germanene.
The adjustable gap $\Delta_z = E_z d$, where $2 d$ is the separation between the two sublattices situated in different  planes,
can be tuned by applying an electric field $E_z$.
The function $f(\varepsilon) =  |\varepsilon|/(\pi  \hbar ^{2} v_{F}^{2})$ is twice smaller than one for graphene, because
the first transition in Eq.~(\ref{DOS-general}) with
$i=1$ corresponds to $\eta = \sigma = \pm$ with $\Delta_1 = | \Delta _{\text{SO}} - \Delta _{z} |$ and
the second one with $i=2$ corresponds to $\eta = -\sigma = \pm$ with $\Delta_2 = |\Delta _{z}+ \Delta _{\text{SO}} |$.

Since the DOS is a symmetric function, instead of the total density of electrons it is convenient to operate with
the difference between the densities of electrons and holes  given by (see~\cite{Tsaran2017}, see the Methids and Appendix~\ref{appendixB} of that paper)
\begin{equation}
\label{number-general}
 n(T,\mu,\Delta_1,\Delta_2,\ldots,\Delta_M)=
\frac{1}{4} \int _{ -\infty }^{\infty }d \varepsilon\, D (\varepsilon) \left [\tanh  \frac{\varepsilon +\mu }{2 T} -
\tanh  \frac{\varepsilon -\mu }{2 T} \right ].
\end{equation}
Clearly, $n (T ,\mu )$ is an odd function of $\mu $ and $n (T ,\mu  =0) =0$.
In Dirac materials the  density $n$  can be controlled by an applied gate voltage. In what follows we
analyse the dependence of $s$ on the chemical potential.

As it was mentioned above, the entropy per particle is directly related to the
temperature derivative of the chemical potential at the fixed density $n$,
see Eq.~(\ref{fullderiv}). 
If the chemical potential is situated between the discontinuity points, $\Delta_i < |\mu| < \Delta_{i+1}$,
and  $T \to 0$,  one obtains
for the first derivative in Eq.~(\ref{fullderiv})
\begin{equation}
\label{dNdT}
\frac{\partial n(T,\mu)}{\partial T}= D^\prime (|\mu|)\frac{\pi^2 T}{3} \,  {\rm sign}(\mu),
\quad \Delta_i > 0.
\end{equation}
On the other hand,  at the discontinuity points $\mu=\pm\Delta_N$ at $T \to 0$, one finds
\begin{equation}
\label{dNdT-disc}
 \left. \frac{\partial n(T,\mu)}{\partial T} \right|_{\mu=\pm\Delta_N}  =  \pm \left[
D(\Delta_N+0)-D(\Delta_N-0)\right]  \int\limits_{0}^\infty \frac{x \, d x}{\cosh^2x}
 = \pm f(\Delta_N)\ln2.
\end{equation}
One can see that a factor of $\ln 2$ originates from the integration of the derivative of the Fermi distribution
(or $\frac{1}{2} \tanh z $) multiplied by the energy.
If $\mu=\pm\Delta_N$ with $N<M$ and $T \to 0$ for the second derivative in Eq.~(\ref{fullderiv}), one obtains
\begin{equation}
\label{dNdmu}
\left.\frac{\partial n(T,\mu)}{\partial\mu}\right|_{\mu=\pm\Delta_N}   = f(\Delta_N)\sum_{i=1}^M\theta(\Delta^2_N -\Delta^2_i)
 =f(\Delta_N)(N-1/2),
\end{equation}
where the first $N-1$ $\theta$ functions contribute $N-1$ and the last one contributes $1/2$.

Thus, we arrive to the conclusion that the entropy per particle in Dirac materials is
\begin{equation}
s(T \to 0,\mu=\pm\Delta_N)=\pm \frac{\ln2}{N-1/2}, \quad N=1,2, \ldots M,
\end{equation}
while for $\Delta_i < |\mu| < \Delta_{i+1}$ it vanishes.
One can see that the behavior of the entropy per particle 
as a function of chemical potential for gapped Dirac systems is analogous to one found in a quasi-2DEG with a
parabolic dispersion considered in Sec.~\ref{2D} and Ref.~\cite{Varlamov2016PRB}. This fact allows us to speculate
that such universal spikes are related rather to the topological changes of the Fermi surface
than to the specific form of the electronic spectrum.

\paragraph{Gapped graphene. --}

In the particular case of a gapped graphene the integral (\ref{number-general}) can be done analytically
\cite{Gorbar2002PRB}
\begin{equation}
\label{number-graphene}
n (T ,\mu,\Delta )=  \frac{2 T^{2}}{\pi \hbar ^{2}v_{F}^{2}}\left[ \frac{\Delta }{T}\ln \frac{%
1+\exp \left( \frac{\mu -\Delta }{T}\right) }{1+\exp \left( -\frac{\mu
+\Delta }{T}\right) }
+\mbox{Li}_{2}\left( -e^{-\frac{\mu +\Delta }{T}%
}\right) -\mbox{Li}_{2}\left( -e^{\frac{\mu -\Delta }{T}}\right) \right],
\end{equation}
where $\mbox{Li}_2 (x)$ is the dilogarithm function. The corresponding derivatives 
are \cite{Tsaran2017} (see also Appendix~\ref{appendixB} of this paper):
\begin{eqnarray}
\label{derivative-T-2nd-a}
\left( \frac{ \partial n }{\partial T }\right)_\mu  & = & \frac{2}{\pi \hbar^2 v_F^2}  \left[
\frac{\Delta}{T} \frac{\mu \sinh (\Delta/T) + \Delta \sinh (\mu/T)}{\cosh (\Delta/T) + \cosh (\mu/T)} 
 + 2 T \mbox{Li}_2 \left(-e^{- \frac{\mu+ \Delta}{T}} \right) - 2 T \mbox{Li}_2 \left(-e^{ \frac{\mu- \Delta}{T}} \right)  \right.
 \nonumber \\
&& + \left. \frac{2 \Delta \mu}{T} - (\mu - 2 \Delta)  \ln \left( 2 \cosh \frac{\mu - \Delta}{2T} \right) -
(\mu + 2 \Delta) \ln \left( 2 \cosh \frac{\mu + \Delta}{2T} \right)
\right], \\
\label{derivative-mu-a}
 \left(\frac{ \partial n }{\partial \mu }\right)_T  &=&  \frac{2}{\pi \hbar
^{2}v_{F}^{2}}\left[ \frac{\Delta }{2}\left( \tanh \frac{\mu -\Delta }{2T}%
-\tanh \frac{\mu +\Delta }{2T}\right) 
+ T\left( \ln \left( 2\cosh \frac{\mu -\Delta }{2T}\right) +\ln
\left( 2\cosh \frac{\mu +\Delta }{2T}\right) \right) \right].
\end{eqnarray}
As one can see, Eq.~(\ref{derivative-mu-a}) is symmetric with respect to the transformation $\mu \to -\mu$
or $\Delta\to-\Delta$. On the other hand  Eq.~(\ref{derivative-T-2nd-a}) is antisymmetric with respect to the swap $\mu \to -\mu$ and
symmetric under the transformation $\Delta\to-\Delta$. The last property is checked using the identity for the dilogarithmic function
\begin{equation}
{\rm Li}_2\left(-\frac{1}{z}\right)=-{\rm Li}_2(-z)-\frac{1}{2}\ln^2(z)-\frac{\pi^2}{6}.
\end{equation}

\paragraph{Low buckled Dirac materials. --}

The density of carriers in silicene can be described with use of the formalism developed above for
graphene by formally representing silicence as a superposition of two gapped graphene layers characterised by different gaps:
$
n  (T ,\mu,\Delta_1, \Delta_2  ) = 1/2 \left[ n (T ,\mu,\Delta_1 ) + n (T ,\mu,\Delta_2 ) \right].
$

\begin{figure}[ht]
\centering
\includegraphics[width=0.99\linewidth]{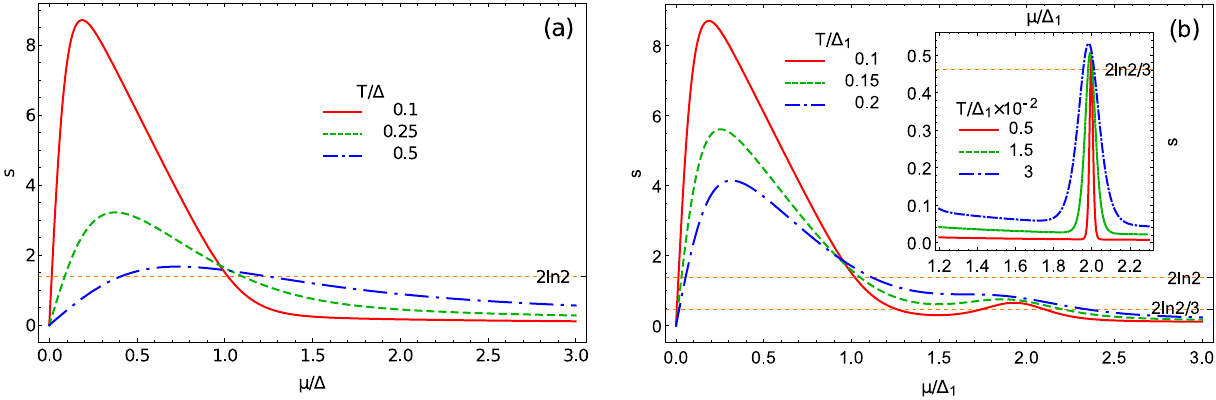}
\caption{ (Color online)  The entropy per electron $s$ vs the chemical potential $\mu >0 $, $s(-\mu) = -s(\mu)$, for three values of temperature.
Left panel: (a): Gapped graphene. The chemical potential $\mu$ is expressed in the units of $\Delta$; the solid (red) $T/\Delta = 0.1$,
dashed (green) $T/\Delta = 0.25$, dash-dotted (blue) $T/\Delta = 0.5$.
Right panel: (b): Silicene. $\mu$ is expressed in the units of the smaller gap $\Delta_1$, the second  gap
$\Delta_2 = 2 \Delta_1$;  the solid (red) $T/\Delta_1 = 0.1$,
dashed (green) $T/\Delta_1 = 0.15$, dash-dotted (blue) $T/\Delta_1 = 0.2$. Zip to the vicinity of
$\mu = \Delta_2$ is shown in the insert:
the solid (red) $T/\Delta_1 = 5 \times 10^{-3}$, dashed (green) $T/\Delta_1 = 1.5 \times 10^{-2}$,
dash-dotted (blue) $T/\Delta_1 = 3 \times 10^{-2}$.  }
\label{fig:1}
\end{figure}

Once the carrier imbalance function, $n(T,\mu,\Delta_1,\Delta_2,\ldots,\Delta_N)$, is found, the entropy per electron
can be calculated using Eq.~(\ref{fullderiv}).
In Fig.~\ref{fig:1} (a) and (b)
we show the dependence $s(\mu)$ for graphene and silicene, respectively,
at three different values of $T$.
Since the entropy per electron is an odd function of $\mu$,
only the region $\mu >0$ is shown.
In the case of silicene we express $\mu$ and $T$
in the units of a smaller gap, $\Delta_1$. The dependence $s(\mu)$ in the vicinity of the second gap,
$\mu = \Delta_2 = 2 \Delta_1$ is shown in the insert of Fig.~\ref{fig:1} (b) to resolve the spike structure
for three temperatures lower than the values on the main plot.

The most prominent feature that we find in Fig.~\ref{fig:1} (a) and (b) is a sharp peak
observed for the chemical potential in the temperature vicinity of the Dirac point,  $|\mu| \sim T$.

Using the low-temperature expansions of the derivatives,
Eqs.~(\ref{derivative-T-2nd-a}) and (\ref{derivative-mu-a}) we consider below the various limiting cases
for the entropy per particle behaviour.
If the chemical potential is within the energy gap but it is not very close to the Dirac point,
$T \ll |\mu|<\Delta$, and $T \ll \Delta-|\mu|$,
the entropy per particle in a gapped graphene is
\begin{equation}
\label{s-mu<Delta}
s(T, \mu,\Delta  )\simeq\mathrm{sign}(\mu )\left[\frac{\Delta-|\mu|}{T}+1+\frac{T}{\Delta+T}
\right].
\end{equation}
Near the Dirac point, $|\mu| \ll T \ll \Delta$,  one finds
\begin{equation}
\label{mu=0}
s(T, \mu ,\Delta  )\simeq \frac{\mu \Delta}{T^2} \left[ 1 + O (e^{-\Delta/T}) \right].
\end{equation}
If the chemical potential crosses the Dirac point at $T=0$, the transition from hole-like to electron-like
carriers is singular. Eqs.~(\ref{s-mu<Delta}) and (\ref{mu=0}) show how the temperature smears it.
The peak inside the gap is mainly due to the specific dependence of the chemical potential on the electron density.
Indeed, since $s= \partial S(T,\mu)/\partial n =  (\partial S(T,\mu)/\partial \mu)(\partial \mu/ \partial n) $,
the dependence $s(\mu)$ is governed by the sharpest function in the product.
The chemical potential grows rapidly  at the small density $n$ and then quickly reaches the value
$|\mu| \simeq \Delta$, where the derivative
$\partial \mu/ \partial n$ becomes small.
The peaked behavior of $s$ may be considered as a smoking gun for the gap opening in gapped Dirac materials.

Near the Lifshitz transition points, $\mu = \pm \Delta$, we observe that the dependences $s(\mu)$ are monotonic
functions, so that these points are not marked by spikes.
This is typical for any system where DOS has just one discontinuity
\cite{Varlamov2016PRB}.
Nevertheless, the entropy per particle quantization rule for graphene
$s(\mu = \pm \Delta) = \pm 2 \ln2$ is fulfilled.
One can see that in both panels of Fig.~\ref{fig:1}, at low temperatures all  curves cross each other near this point.
The corresponding value $s= 2 \ln 2$ is shown  by the dotted line.
This numerical result can be confirmed analytically.
For $T \ll \Delta$ we obtain
\begin{equation}
\label{2ln2-T}
s(T,\mu=\Delta,\Delta)=2\ln2+\frac{\pi^2-12\ln^22}{3}\frac{T}{\Delta}+
O\left(T^2\right).
\end{equation}

Now we briefly discuss the effect of broadening of the energy levels due to
the scattering from  static defects.
Let us smear the DOS function (\ref{DOS-general})
by convolving it with the Lorentzian, $\gamma/[\pi(\omega^2+\gamma^2)]$,
where $\gamma$ is the scattering rate.
In the regime $\gamma\ll T\ll\Delta$ one finds
\begin{equation}
\label{2ln2-dirty}
s(T, \mu = \Delta, \Delta)=2\ln2 \left[1-\frac{\gamma}{T}\left(\frac{1}{\pi\ln2}
+\frac{T}{\Delta}\right)\right].
\end{equation}
Eq.~(\ref{2ln2-dirty}) shows that the universality of the low temperature entropy per particle
is broken by the disorder if the mean free path becomes comparable with the thermal diffusion length.

The case $\Delta=0$ deserves a special attention. In this limit, Eq.~(\ref{number-graphene})
acquires a simple form (see~\cite{Tsaran2017}, Methods and Appendix~\ref{appendixB} of that paper). 
For the entropy per particle one finds
\begin{equation}
\label{s(T,0,mu)}
s(T,\mu,0)=\left\{%
\begin{array}{ll}
\frac{\mu}{T}\left(1-\frac{\mu^2}{T^2}\frac{1}{6\ln2}\right), & |\mu| \ll T,
  \\
\frac{\pi^2}{3}\frac{T}{\mu},& T\ll |\mu| . 
\end{array}%
\right.
\end{equation}
It is important to note that if the second line of Eq.~(\ref{s(T,0,mu)})
is multiplied by the factor $k_B/e$ it yields the Seebeck coefficient for a free electron gas \cite{Abrikosov.book}.
Moreover, the general expression for $s = -\partial \mu/\partial T$, Eq.~(\ref{derivative-Delta=0}),
reproduces the thermal power $S$ that can be extracted  from the results based on the Kubo formalism \cite{Sharapov2012PRB}
that validates  the thermodynamic approach of \cite{Varlamov2013EPL}.

The presence of the second gap in silicene and similar materials, $\Delta_2 > \Delta_1$, results in the appearance of the
peak in $s(\mu) \approx \pm  2 \ln2/3$ near the point $\mu = \pm \Delta_2$, as seen in Fig.~\ref{fig:1}~(b).
The corresponding value $s= 2 \ln2/3$ is shown by the dotted line.
This peak can be considered as a signature of the second Lifshitz transition which occurs if $\mu$
crosses $\Delta_2$. Indeed, as we have shown in the previous Section (see also  in \cite{Varlamov2016PRB}) for the quasi-2DEG
the peak structure in $s(\mu)$ develops only if the number of discontinuities in the DOS, $N \geq 2$.
We conclude that perspective Dirac materials, such as silicene and germanene, where the spin orbit interaction plays a very important role
allow the simplest realization of the $N = 2$ case with two discontinuities on  both electron
and hole sides of the total DOS.

\begin{figure}[ht]
\centering
\includegraphics[width=0.99\linewidth]{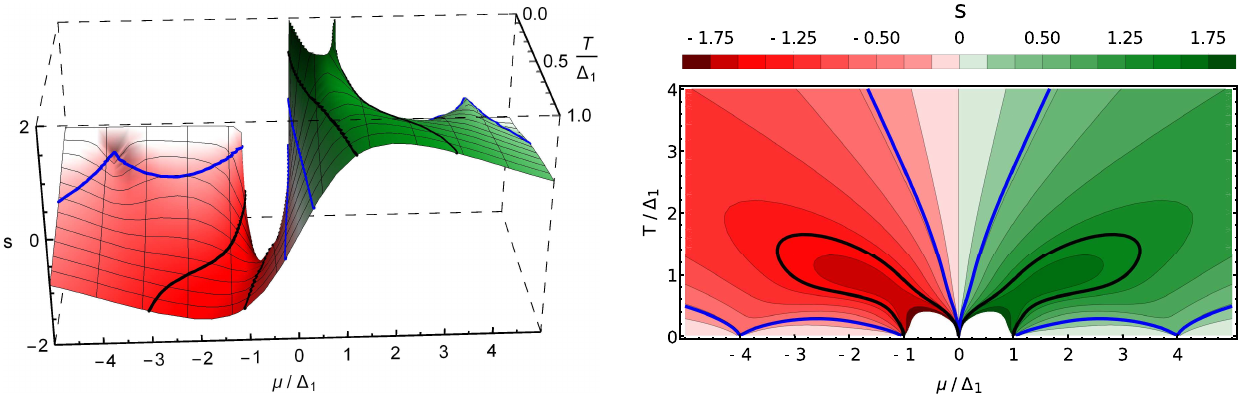}
\caption{{ The entropy per electron $s$
as functions of the chemical potential $\mu$ and temperature $T$ in the units of $\Delta_1$.} The gap $\Delta_ 2 = 4\Delta_1 $.
Left panel:  3D plot.  Right panel:  Contour plot. }
\label{fig:2}
\end{figure}

Figure~\ref{fig:2} shows the 3D and density plots of $s$ as a function of
$\mu/\Delta_1$ and $T/\Delta_1$.
To be specific, we assumed that $\Delta_1$ is the smallest of the gaps and chose $\Delta_2 = 4 \Delta_1$.
The black and blue lines correspond to the contours of constant values
$s = \pm 2 \ln 2$ and $s = \pm 2 \ln 2 /3$, respectively.
The range of $s$ in the 3D plot is restricted to $-2 \leq s \leq 2$, so that only the peaks at $\mu =\pm \Delta_2$ can be seen.

A more careful examination of Fig.~\ref{fig:1}~(b) shows that the peak occurring near  $\mu = \Delta_2$
is somewhat shifted to smaller than $\Delta_2$ values of $\mu$. Looking at  Fig.~\ref{fig:1}~(b) and its insert
one can trace  how the position of this peak moves towards the point $(\mu = \Delta_2, T =0)$
as the temperature decreases. In Fig.~\ref{fig:1}~(a) the increase of its height can be seen.
Close to this point ($T \ll \Delta_2$) we obtain analytically
\begin{equation}
\label{2ln2-T-2nd}
s(T,\mu=\pm\Delta_2)=\pm\left[\frac{2\ln2}{3}+\frac{\pi^2-4\ln^22}{9}\frac{T}{\Delta_2}\right].
\end{equation}
In what concerns the behaviour of the entropy per particle in the vicinity of the smallest gap in silicene, $\Delta_1$,
it is described by  Eq.~(\ref{2ln2-T}) with  $\Delta$ replaced by $\Delta_1$.

\begin{figure}[ht]
\centering
\includegraphics[width=0.99\linewidth]{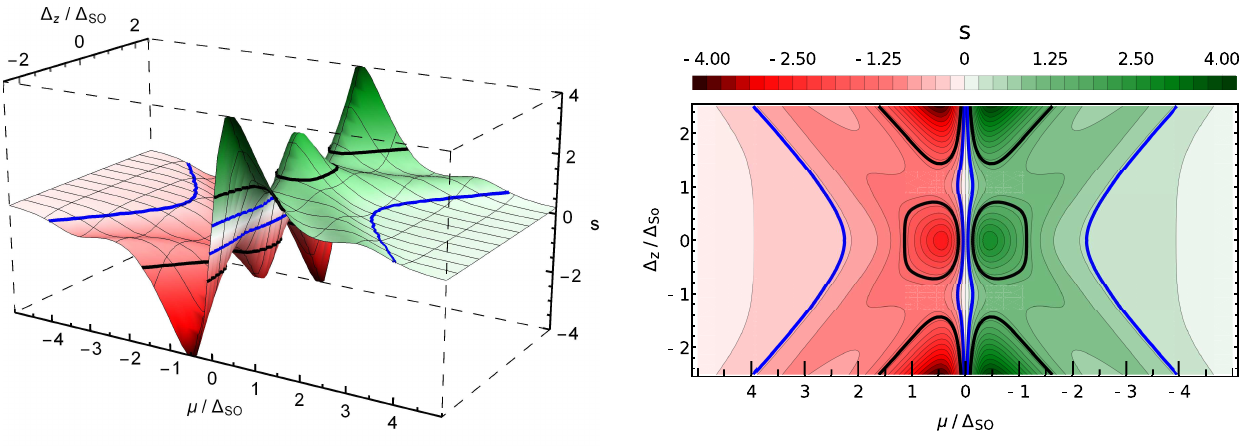}
\caption{{ The entropy per electron $s$
as functions of the chemical potential $\mu$ and $\Delta_z$  in the units of $\Delta_{\mathrm{SO}}$. The temperature $T =0.3 \Delta_{\mathrm{SO}}$.}
Left panel:  3D plot.  Right panel:  Contour plot. }
\label{fig:3}
\end{figure}

Recent successes in fabrication of silicene field-effect transistors \cite{Tao2015NatNano}
offer the opportunity of a direct measurement of the entropy per particle in this promissing 2D material.
In the prospective experiment, a double gate structure would be needed that enables one to tune $\mu$ and $\Delta_z$
independently. Such a situation is modelled in Fig.~\ref{fig:3}, where we  show the 3D and density  plots of $s$ as a function of
$\mu/\Delta_{\mathrm{SO}}$ and $\Delta_z/\Delta_{\mathrm{SO}}$.
As in Fig.~\ref{fig:2}, the black and blue lines are constant value lines with
$s = \pm 2 \ln 2$ and $s = \pm 2 \ln 2 /3$, respectively.
The points $\Delta_z = \pm \Delta_{\mathrm{SO}}$ correspond to the case where
$\Delta_1 =0$ and $\Delta_2 = 2 \Delta_{\mathrm{SO}}$ or $\Delta_1 = 2 \Delta_{\mathrm{SO}}$ and $\Delta_2 =0$, so that
the system experiences a transition from a two-gap to a one-gap spectrum.
For $|\Delta_z|  < \Delta_{\mathrm{SO}}$ the system is a topological insulator
and for $|\Delta_z|  > \Delta_{\mathrm{SO}}$ it is a trivial (band) insulator.

\paragraph{Discussion.  --}

We presented in this Section the analytical expressions for the entropy per particle in a wide energy range for various Dirac materials.
Based on these expressions we have predicted the characteristic spikes of the entropy per particle at the
Lifshitz topological transition points in several 2D Dirac systems.
The magnitude of spikes is quantized at low temperatures and it is independent of material parameters.
The quantized spikes are expected to occur in silicene and germanene. They can also be found in the
gapped graphene in the presence of Zeeman splitting and in quasi two-dimensional Dirac and Weyl materials.
Note that the quantization of entropy and spikes of the same origin occur in a 2DEG  in
the presence of Zeeman splitting~\cite{Pudalov} (see \cite{Tsaran2017}, Methods and Appendix~\ref{appendixB} of this paper).

Our results are based on the assumption that the function $f(E)$ in the
DOS (\ref{DOS-general}) is continuous. Although this assumption is quite
general, it is not fulfilled, for example, in  a bilayer graphene.
The overall behavior of the entropy per electron $\partial S/\partial n$ as a function of the electronic chemical potential
may be used as a tool for characterization of the electronic dispersion in novel crystal structures.
The crucial point is that $\partial S/\partial n$ is related to the temperature derivative
$\partial \mu/\partial T$ via 
the relation (\ref{fullderiv}).  The last value, as it has been mentioned in Introduction, can be directly measured using the experimental
approach  developed in \cite{ Pudalov}. It appears that this technique has three orders of
magnitude higher resolution than the other methods and thus it can be very helpful in probing interaction effects
in 2D electron systems.
The measurements of the entropy per particle can also be used to study
the effect of interactions on the DOS in graphene, because
the renormalization of the Fermi velocity due to electron-electron interactions \cite{Elias2011NatPhys}
modifies the function $s (n)$.

Another interesting point to address is that the motion of electrons in
graphene can become hydrodynamic when the frequency of electron-electron
collisions is much larger than the rates of both, electron-phonon and
electron-impurity scattering  (see Refs.~\cite{Narozhny17,Lucas18} and references therein). The
discussion of these effects goes beyond the scope of the present review.
We mention however that this regime is expected to be accessible in the
ultra-pure materials and in an intermediate temperature range. The
hydrodynamic effects are observed via transport measurements for strongly
interacting gapless quasiparticles in graphene at the charge-neutrality
point~\cite{Narozhny17,Lucas18}. In our studies of Dirac materials including graphene we deal
with thermodynamics of free gapped quasiparticles in these materials where
the interaction effects are partially taken into account
phenomenologically through gaps, quasiparticle velocities and
quasiparticle widths. 

It would be also interesting to address the role of electron-phonon interaction in low-dimensional materials, which may not be taken into account by effective renormalization of relevant parameters. This issue also falls outside the present review.
As we mentioned above, deviations from predicted
effects might indicate a range
of temperatures and particle densities where electron-electron and
electron-phonon interactions become appreciable.


\section{Entropy per   particle in transition metal dichalcogenides}\label{dichalcogenides}

Layered transition-metal dichalcogenides (TMDCs)  represent another class
of materials whose monoatomic layers are being studied experimentally. We show in this Section 
based on~\cite{Shubnyi2017}
that similar effects to those discussed in the previous Sections may be observed also in TMDCs.
Single layer TMDCs  with the composition MX$_2$
(where M = Mo, W is a transition metal, and X = S, Se, Te is a chalcogen atom)
are truly two-dimensional (2D) semiconductors with large band gaps ranging within {1}{2}~{eV} (see, e.g., Refs.~\cite{Chhowalla2013NatChem,Kormanyos20152DMat}).
Consequently, one may expect that in these materials the peaks in entropy per particle can be seen at much higher temperatures than in silicene and germanene.
  
We begin with presenting the model which describes the single layer TMDCs. Since the full description of strained TMDCs is very complicated, the effect of a uniform uniaxial strain is taken into account only via scalar potential  spin-independent parts. 
Then we discuss the DOS and present an analytical expression for the entropy per particle
in TMDCs.  The results for the obtained behavior of  the entropy per particle are discussed 
and conclusions are given in the remaining part of this Section. 

\paragraph{Hamiltonian. --}
\label{sec:model}

The low-energy excitations in monolayer TMDCs can be described by the following model Hamiltonian density
\cite{Capellutti2013PRB,Rostami2013Rostami,Liu2013PRB,Ridolfi2015JPCM,Kormanyos20152DMat}
\begin{equation}
 H  = \sum_{\tau = \pm 1} H_\tau, \quad
 H_\tau  = H_{D}^{\tau} + H_{2}
\end{equation}
where $\tau=\pm 1$ is the valley index,
$H_{D}^{\tau}$ is linear in momentum in the Dirac-like part of the band-structure \cite{Xiao20012PRL}
and $H_2$ is the quadratic part of the Hamiltonian. The Dirac Hamiltonian contains free massive Dirac fermion kinetic energy term, $H_{0}^\tau $,
and the spin-orbit term $H_{SO}^\tau$, $H_{D}^{\tau} = H_{0}^\tau + H_{SO}^\tau$.
The first term is
\begin{equation}
\label{H_0}
H_{0}^\tau = \hbar v_F(\tau k_x\sigma_x+k_y\sigma_y)+\frac{\Delta}{2}\sigma_z -\mu \sigma_0,
\end{equation}
$\pmb{\sigma}$ are the Pauli matrices acting in the $2 \times 2$ ``band'' space,
$\sigma_0$ is the unit matrix, the Fermi velocity
$v_F = a t/\hbar \sim  0.5 \times 10^6$~{m/s} with $t$ being the effective hopping integral
and $a$ is the lattice constant, the major band gap $\Delta \sim \text{ range {1}{2}~{eV}}$.
The inversion symmetry breaking results in the
the spin-orbit part of the Hamiltonian
\begin{equation}
\label{H_SO}
H_{SO}^\tau=\lambda_v\tau\frac{\sigma_0-\sigma_z}{2}s_z + \lambda_c\tau\frac{\sigma_0+\sigma_z}{2}s_z,
\end{equation}
where $s_z$ is the Pauli matrix for spin,
$2\lambda_v \sim \text{150 - 500}$~{meV} is the spin splitting at the valence band top
caused by the spin orbit coupling, $2 \lambda_c$ is the spin splitting  at
at the conduction band bottom. The DFT calculations \cite{Kormanyos20152DMat} show that the
absolute value $2\lambda_v \gg  |2\lambda_c |\sim \text{{3} - {50}}$~{meV} and the sign
of $\lambda_c$ depends on the compound, $\lambda_c >0$  for MoX$_2$ and $\lambda_c < 0$ for WX$_2$ compounds.

The quadratic part of the Hamiltonian, $H_2$, contains the following diagonal terms
\begin{equation}
\label{H_2}
H_2 =\frac{\hbar^2 k^2}{4 m_e} (\alpha \sigma_0 + \beta \sigma_{z}),
\end{equation}
where $m_e$ is the free electron mass, and $\alpha \neq \beta$ are constants of the order of $1$.
Finally, as discussed in \cite{Capellutti2013PRB,Rostami2013Rostami,Liu2013PRB,Ridolfi2015JPCM,Kormanyos20152DMat}
more accurate approximations also include the trigonal warping terms.

The spin-up and spin-down components are completely decoupled,
thus the spin index $\sigma=\pm1$ is a good quantum number.
Neglecting the quadratic term (\ref{H_2}) we obtain
the dispersion dependencies for conduction and valence bands
\begin{equation}
\label{TMD-spectrum}
\varepsilon_{c,v}(k)  =  \frac{\lambda_v+\lambda_c}{2}\tau \sigma \\
 \pm\sqrt{ \hbar^2 v_F^2 k^2+ (\Delta-(\lambda_v-\lambda_c)\tau \sigma)^2/4}.
\end{equation}
This electronic spectrum closely resembles one of massive  fermions
in low-buckled Dirac materials described by Eq.~(\ref{silicene-spectrum}). This is valid with the oteworthy exception to the first
valley- and spin-dependent term in Eq.~(\ref{TMD-spectrum}).
In the first order approximation, one can neglect the conduction band splitting and take $\lambda_c =0$ in order to formulate the
simplest model \cite{Xiao20012PRL}, where  the conduction band remains spin degenerate at $\mathbf{K}$ and $\mathbf{K}^\prime$
points and it has a small spin splitting quadratic in $\mathbf{k}$, whereas the valence band is completely split,
\begin{equation}
\label{TMD-spectrum-simple}
\varepsilon_{c,v}(k)=\frac{\lambda_v}{2}\tau  \sigma \pm\sqrt{\hbar^2 v_F^2 k^2+(\Delta-\lambda_v\tau \sigma)^2/4} .
\end{equation}

Single layer TMDCs  can sustain deformations higher than
10\%  \cite{Bertolazzi2011ACSNano,Castellanos2012AdvMat}. The experimental possibility
to tune the band gap with strain has been proven for MoS$_2$ in
\cite{Conley2013NanoLett,Hui2013ACSNano,Castellanos-Gomez2013NanoLett,Zhu2013PRB}
and in WS$_2$ \cite{Wang2015NanoRes,Voiry2013NatMat,Georgiou2013NatNano}.
The full description of strained TMDCs is much more involved than that of graphene
as it involves five different fictitious gauge fields as well as scalar potentials
entering spin-independent and spin-dependent parts
\cite{Rostami2015PRB}. Below we restrict ourselves to the qualitative description of the strain
effect on the properties of TMDCs, and we consider only the scalar potential term in the
spin-independent Hamiltonian (\ref{H_0}), viz.
\begin{equation}
\label{strain}
H_{\mathrm{str}} = \frac{D_{+}( \hat{\varepsilon}) + D_{-} ( \hat{\varepsilon})}{2} \sigma_0 +
\frac{D_{+} ( \hat{\epsilon}) - D_{-} ( \hat{\epsilon})}{2} \sigma_3,
\end{equation}
where $ \hat{\epsilon}$ is the strain tensor.
The explicit expressions for the diagonal terms $D_{\pm}$ are provided in \cite{Rostami2015PRB} and here
we only keep the linear in strain contributions neglecting the higher order terms
\begin{equation}
\label{D}
D_{\pm} = \alpha_2^{\pm} (\epsilon_{xx} + \epsilon_{yy}) ,
\end{equation}
with $\alpha_2^{+} =  {-3.07}$~{eV} and  $\alpha_2^{-} =  {-1.36}$~{eV}.
The corresponding parameters 
for the spin-dependent part are smaller by the three orders of magnitude, so that
the corresponding term in the Hamiltonian can be safely neglected.
Assuming that the strain is a uniform uniaxial one, we can express $D_{\pm}$ via  $\epsilon \equiv \epsilon_{xx}$
($\epsilon >0$ for tensile strain) and the
Poisson ratio $\nu$~\cite{Landau.book}  as $D_{\pm}= \alpha_2^{\pm} \epsilon (1- \nu)$.
Thus, in the present toy model the effect of strain is reduced to the renormalization of
the chemical potential,
\begin{equation}
\label{mu-strain}
\mu \to \mu - \epsilon (1-\nu)(\alpha_2^{+} + \alpha_2^{-})/2
\end{equation}
and of the gap
\begin{equation}
\label{gap-strain}
\Delta \to \Delta + \upsilon (1-\nu)(\alpha_2^{+} - \alpha_2^{-}).
\end{equation}
Setting $\nu =0$ one can estimate that 1\% tensile strain shifts $\mu$ by 22~meV and
$\Delta$ by {17}~meV, respectively.

\paragraph {Asymmetric DOS and modification of entropy quantization. -- }
\label{sec:entropy}

As it has been mentioned above several times, the entropy per particle is directly related to the
temperature derivative of the chemical potential at the fixed density $n$
through Eqs.~(\ref{fullderiv}) and (\ref{ngen}).

Note that in the presence of the electron-hole symmetry
it is convenient to operate with the difference $n$ between the densities of electrons and holes instead of the total
density of electrons, as it is usually done for graphene \cite{Tsaran2017}.
One can show that in a close analogy with graphene and low-buckled Dirac materials
the DOS for TMDCs described by the approximate spectrum
(\ref{TMD-spectrum}) is
\begin{equation}
\label{TMD-DOS}
D(\varepsilon)=\frac{1}{\pi(\hbar v_F)^2}\sum_{i = \pm 1}
\left| \varepsilon- E_i \right|
\theta\left[(\varepsilon- E_i)^2- \Delta_i^2\right].
\end{equation}
Here we denoted
$E_i = i(\lambda_v + \lambda_c) /2 $ and
$\Delta_i = [\Delta-i(\lambda_v - \lambda_c)]/2$ with
$i=+1$ corresponding to $\tau = \sigma = \pm 1$ and
$i=-1$ corresponding to $\tau = -\sigma = \pm 1$.

Obviously,  for $\lambda_c =0$  the resulting DOS corresponds to the energy spectrum
(\ref{TMD-spectrum-simple}). The DOS (\ref{TMD-DOS}) differs from the one described by the
equation (\ref{DOS-general}) due to the presence of the energy shift, $E_i$, in the absolute value
and in the argument of the $\theta$-function.
As a consequence, the quantization of the entropy per particle, $s = \pm 2 \ln2/3$,
obtained in \cite{Tsaran2017}  for the low-buckled Dirac materials
does not occur in TMDCs.

\begin{figure}[!h]
\centering
\includegraphics[width=8cm]{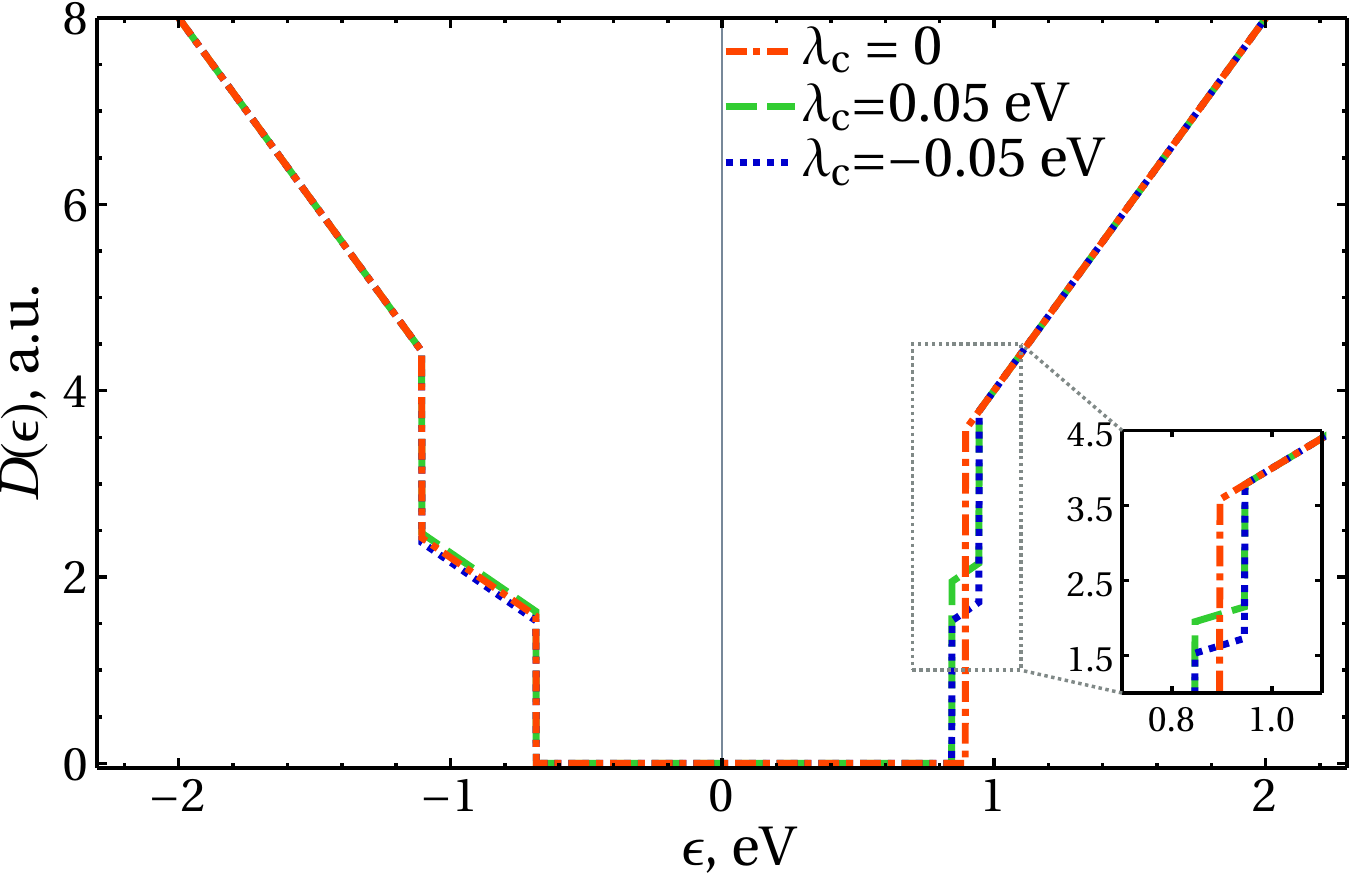}
\caption{(Color online) The DOS, $D (E)$, in arbitrary units versus energy in eV.
The parameters are $\Delta =  {1.79}$~{eV}, $2\lambda_v=  {0.43}$~{eV}. The dashed-dotted (red) line $\lambda_c=0$,
long dashed (green) $\lambda_c=  {0.05}$~{eV}, dotted (blue) $\lambda_c=  {-0.05}$~{eV}.
} \label{fig:1a}
\end{figure}

The behavior of the DOS given by Eq.~(\ref{TMD-DOS}) is illustrated in  Fig.~\ref{fig:1a}.
To be specific, we took the values $\Delta =  {1.79}$~{eV}, $2\lambda_v=  {0.43}$~{eV} corresponding to
the compound WS$_2$.
The constant $2\lambda_c$  for WS$_2$  is $ {-0.03}{eV}$ \cite{Kormanyos20152DMat}.
In order to demonstrate the role of this parameter we have chosen the larger value of $\lambda_c$.
Furthermore, we have considered three possible cases: $\lambda_c =0$
is shown by  the dash-dotted (red) line, long dashed (green) line is for $\lambda_c =  {0.05}$~{eV},
dotted (blue) line is for  $\lambda_c =  {-0.05}$~{eV}.
Note that, in general, {\it ab initio} density functional theory calculations \cite{Kormanyos20152DMat} predict that
$\lambda_c >0$ and  $\lambda_c <0$ correspond to MoX$_2$ and WX$_2$  compounds.
Going from the negative to positive energies we observe the first discontinuity of the DOS
at $E_{-1}^{-} = - \Delta/2 - \lambda_v =  {-1.11}$~{eV}. It linearly goes down until the second discontinuity
that occurs at  $E_{1}^{-} = - \Delta/2 + \lambda_v =  {-0.68}$~{eV}. Their positions are independent of the value of $\lambda_c$.
The DOS is zero inside the gap between $E_{1}^{-}$ and $E_{-1}^{+} =  \Delta/2 - \lambda_c$.
Then it increases linearly until it reaches the discontinuity at the energy
$E_{1}^{+} =  \Delta/2 + \lambda_c$.
Obviously, for $\lambda_c =0$ the last two discontinuities  become degenerate $E_{-1}^{+} = E_{1}^{+} =  {0.895}$~{eV}.
For a finite $\lambda_c$, their ordering depends on the sign of $\lambda_c$.

The peculiarities of DOS  in  TMDCs beyond the Dirac approximations are discussed
in \cite{Scholz2013PRB,Iurov2017}. The quadratic part of the Hamiltonian  (\ref{H_2}) results in the
curving of the electronic dispersion seen in Fig.~\ref{fig:1a}.
Such curving is not essential and it  does not affect the discontinuous character of the DOS function
that is responsible for the peaks in $s(\mu)$.

An advantage of the linearized approximation is that it resembles the case of gapped
graphene and it allows to obtain rather simple analytical results, therefore. For example, one can derive the analytical expression
for the particle density (carrier imbalance)  \cite{Gorbar2002PRB} and find the derivative $\partial \mu/\partial T$
using Eq.~(\ref{fullderiv}). Its generalization  for the low-buckled Dirac materials was made in
\cite{Tsaran2017} (see also \cite{Iurov2017b}). The expression for the particle density in
TMDCs beyond the Dirac approximation is discussed in \cite{Iurov2017}, but it is not very practical for
obtaining the derivative $\partial \mu/\partial T$.

Differentiating Eq.~(\ref{number}) for the electron concentration
\begin{equation}
\label{number}
n_{\text{tot}} (T,\mu)  =\int _{-\infty}^{\infty }d \varepsilon \,  D (\epsilon ) f_{\rm{FD}} \left(\frac{\varepsilon -\mu }{T} \right),
\end{equation}
 where $f_{\rm{FD}} ( x ) = 1/[\exp(x)+1 ]$ is the Fermi-Dirac distribution function
dependent on $T$ and $\mu$. Here we are shifting the variable of integration
$E \to E + E_i$ for each term in the DOS (\ref{TMD-DOS}).
Eventually, we obtain
\begin{equation}
\label{derivative-T-gen}
\left( \frac{\partial n_{\text{tot}}}{\partial T} \right)_\mu
= \int _{-\infty}^{\infty } \frac{d \varepsilon \,  (\varepsilon - \mu)  D (\varepsilon )  }{4 T^2 \cosh^2 \frac{ \varepsilon- \mu}{2T}}
= \frac{1}{2}\sum_{i = \pm 1}  \frac{\partial n (\mu_i,\Delta_i,T)}{\partial T}
\end{equation}
and
\begin{equation}
\label{derivative-mu-gen}
\left( \frac{\partial n_{\text {tot}}}{\partial \mu }\right) _{T}
= \int _{-\infty}^{\infty } \frac{d \varepsilon \,  D (\varepsilon )}{4 T \cosh^2 \frac{\varepsilon - \mu}{2T}}
=  \frac{1}{2} \sum_{i = \pm 1} \frac{\partial n (\mu_i,\Delta_i,T)}{\partial \mu},
\end{equation}
where $\mu_i=\mu-E_i$ is the shifted chemical potential and the derivatives are given by Eqs.~(\ref{derivative-T-2nd-a})  and (\ref{derivative-mu-a}). 
\begin{figure}[!h]
\centering{
\includegraphics[width=10cm]{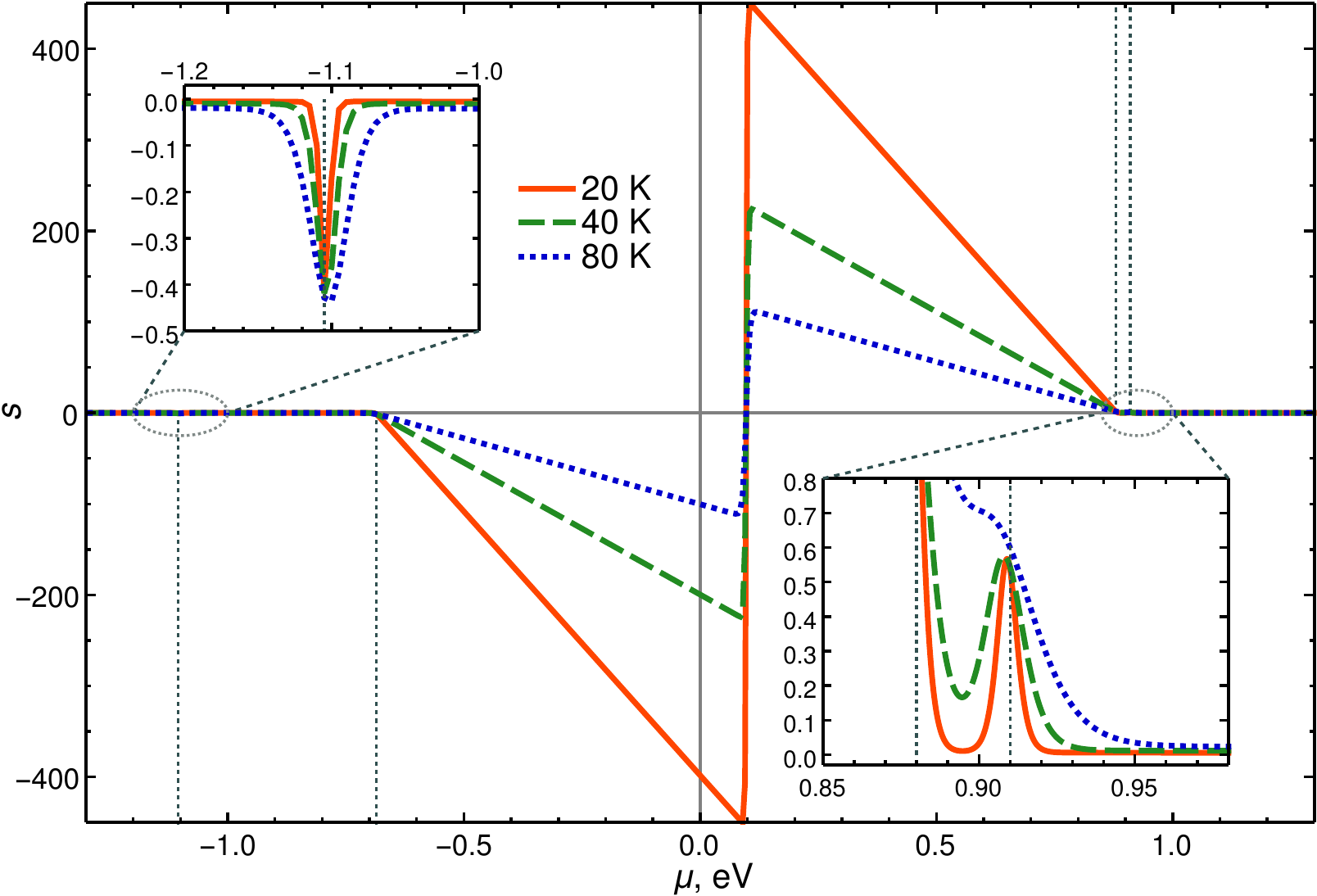}}
\caption{(Color online) The entropy per electron $s$ vs the chemical potential $\mu$ in eV for three values of temperature.
The parameters are $\Delta =  {1.79}$~{eV}, $2\lambda_v=  {0.43}$~{eV} and $\lambda_c= - {0.015}$~{eV}.
 }
\label{fig:2a}
\end{figure}
\paragraph {Results. -- }
Based on obtained results one can investigate
the dependence $s(\mu)$ for the different cases.
Figure~\ref{fig:2a} is computed for the material parameters $\Delta$, $\lambda_v$ and $\lambda_c$
chosen for WS$_2$ compound. The dependence $s(\mu)$ is shown for three values of the temperature:
the solid (red) line is for $T = {20}$~{K}, the dashed (green) line is for $T = {40}$~{K} and
the dotted (blue) line is for $T = {80}$~{K}. The vertical lines correspond to the values of the chemical potential
$\mu = \epsilon_{-1}^{-}, \epsilon_{1}^{-}, \epsilon_{1}^{+}, \epsilon_{-1}^{+}$
that correspond to the discontinuities of the DOS.
Comparing Fig.~\ref{fig:2a} with the results presented in \cite{Tsaran2017}, one can
see that the overall shape of $s(\mu)$ is similar for TMDCs and low-buckled Dirac materials,
although the details are different. For example, inside the gap for $\mu \in [E_1^{-}, E_1^{+}]$
the dependence of $s$ on the chemical potential exhibits a high-amplitude dip-and-peak structure in the
temperature vicinity of the point $\mu = (\lambda_v +\lambda_c)/2$. (The value $i=1$ corresponds to the smaller gap in
Eq.~(\ref{TMD-DOS})). This feature is
even more pronounced and sharp in TMDCs than in the other materials due to the larger ratio $\Delta/T$.
However in the low-buckled Dirac materials this structure is present in the temperature vicinity of the Dirac point,
$\mu =0$, because the whole dependence $s(\mu)$ is an antisymmetric function of $\mu$ in silicene and germanene. This is obviously not the case of  TMDCs.
As discussed in \cite{Tsaran2017} the peak inside the gap is mainly due to the specific dependence
of the chemical potential on the electron density.

The presence of the second larger gap, $\Delta_2 > \Delta_1$, in silicene and similar materials
results in the emergence of the peak in $s(\mu)$ near the points $\mu = \pm \Delta_2$. Similarly,
the discontinuities of the DOS given by Eq.~(\ref{TMD-DOS}) at $\mu = E_{-1}^{-},  E_{-1}^{+}$  associated with
a larger gap $i=2$ also result in the peaks in $s(\mu)$. They are shown in the insert in Fig.~\ref{fig:2a}.
Note that these features are of much weaker amplitude. As explained above, the value of $s$ at the peaks in TMDCs in the low temperature limit is not
equal to the quantized value $\pm 2 \ln 2/3$ expected for the low-buckled Dirac materials \cite{Tsaran2017}.
It is essential that both peaks can still be seen at rather high temperatures. The peak on the right starts to smear
at $T =  {80}$~{K}, while the peak on the left can still be seen at this and higher temperatures.

\begin{figure}[!h]
\centering{
\includegraphics[width=10cm]{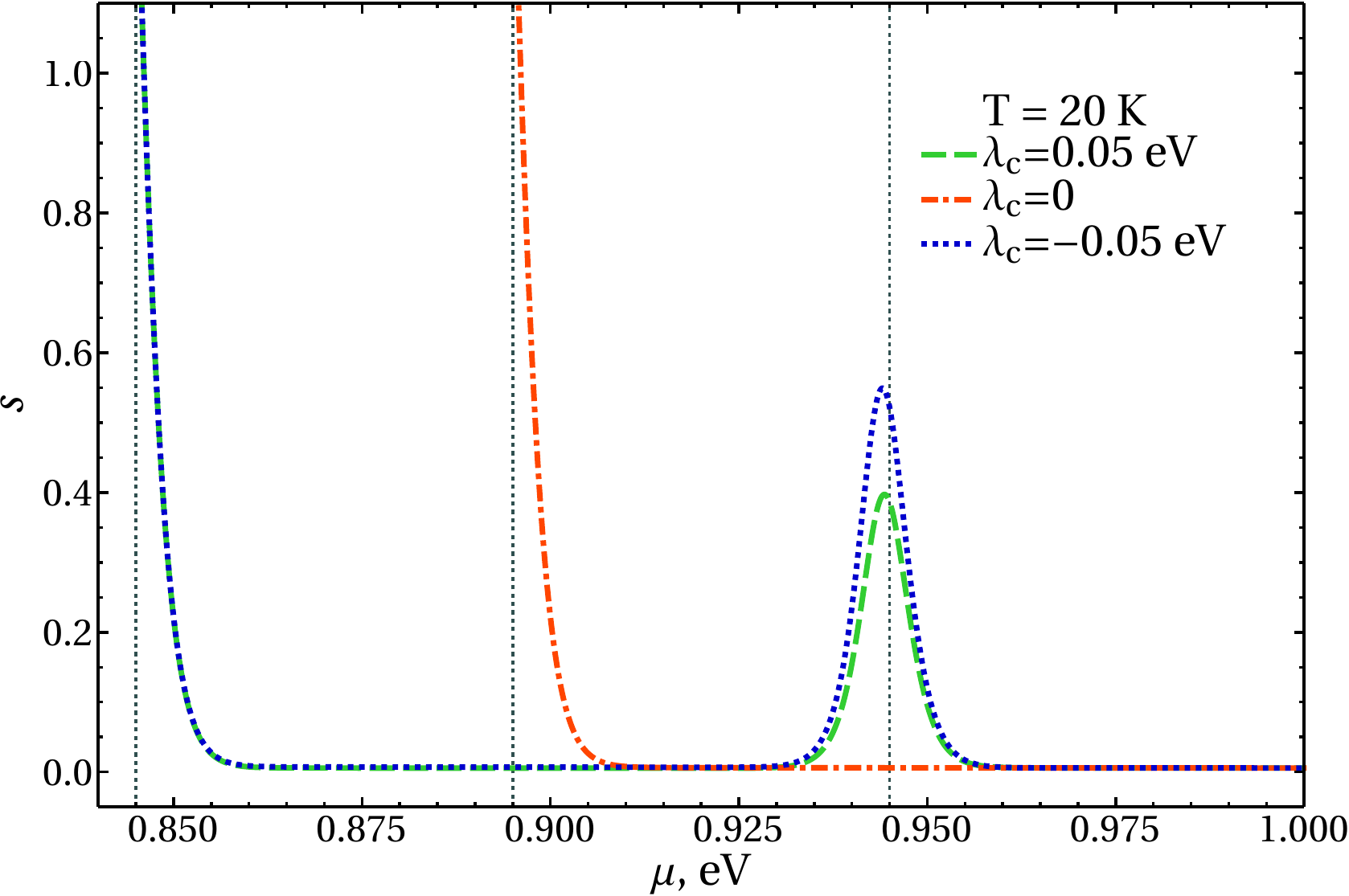}}
\caption{(Color online) The entropy per electron $s$ vs the chemical potential $\mu$ in  eV for three values of $\lambda_c = 0, \mp  {0.05}$~{eV}.
The parameters are $\Delta =  {1.79}$~{eV}, $2\lambda_v=  {0.43}$~{eV} and the temperature
$T = {20}$~{K}.
} \label{fig:3a}
\end{figure}

It is shown in Fig.~\ref{fig:1a} that for $\lambda_c =0$ the two discontinuities of the DOS merge at
$\mu = E_{-1}^{+} = E_{1}^{+}$. Then the positive peak in $s(\mu)$ disappears as the
dash-dotted (red) curve shows in Fig.~\ref{fig:3a}.
As in Fig.~\ref{fig:2a} the vertical lines correspond to the singularities of the DOS.
There is only one singularity for the dash-dotted (red) line at $\mu = E_{-1}^{+} = E_{1}^{+} =  {0.895}$~{eV}.
For nonzero $\lambda_c$ there are two singularities shifted from this point to the left and right by $\mp |\lambda_c| =  {0.05}$~{eV}.
In this case, the peak at the larger energy $\mu = \Delta/2 + |\lambda_c|$ is restored as the
dotted (blue) line shows for $\lambda_c < 0$ and the long dashed (green) line shows for  $\lambda_c >0 $.
\paragraph {The effect of uniform uniaxial strain. -- }
Finally we discuss here how a uniform uniaxial strain would affect the results shown in Fig.~\ref{fig:2a}.
We use Eqs.~(\ref{mu-strain}) and (\ref{gap-strain}) to model the dependence of the chemical potential and the energy gap $\Delta$
on the strain, respectively.
\begin{figure}[!h]
\centering{
\includegraphics[width=10cm]{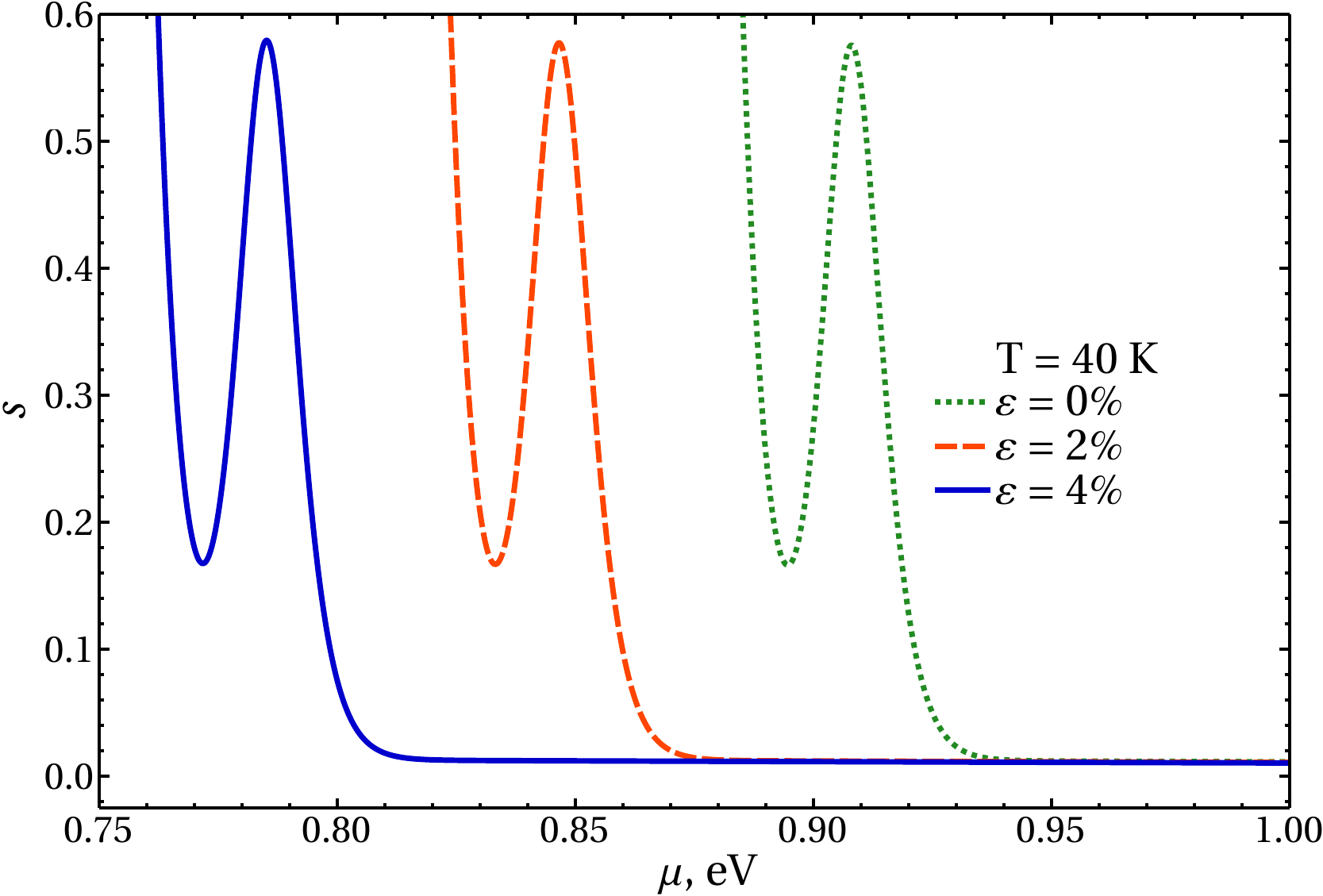}}
\caption{(Color online) The entropy per electron $s$ vs the chemical potential $\mu$ in eV for three values of
strain. The parameters are
$\Delta =  {1.79}$~{eV}, $2\lambda_v=  {0.43}$~{eV},  $\lambda_c= - {0.015}$~{eV},
$\alpha_2^+ =  {-3.07}$~{eV},  $\alpha_2^- =  {-1.36}$~{eV} and the temperature
$T = {40}$~{K}. }
\label{fig:4}
\end{figure}
The dependence $s(\mu)$ is shown in Fig.~\ref{fig:4} for three values of the strain:
the dotted (green) line is for $\epsilon=0$,
the dashed (red) line is for $\epsilon =2 \%$ and
the solid (blue) line is for $\epsilon =4 \%$. As expected, the presence of strain results
in the movement of the peaks in $s(\mu)$.


To conclude, in the present Section we had derived a general expression for the entropy per particle  as a function of the chemical potential,
temperature, and gap magnitude for single layer transition metal dichalcogenides subjected to the  uniform uniaxial
strain. The spectrum of quasiparticle excitations of these materials is similar to that of the
low-buckled Dirac materials, as there is the valley- and spin-dependent gap
$\Delta_{\tau \sigma} = [\Delta- \tau \sigma(\lambda_v - \lambda_c)]/2$ in the electronic spectrum.
The difference from the low-buckled Dirac materials is that the whole spectrum of TMDCs is also shifted by a
valley- and spin- dependent constant $E_{\tau \sigma} = \tau \sigma (\lambda_v + \lambda_c) /2 $. This
introduces the hole-electron asymmetry in the band structure of TMDCs and makes the  resulting DOS (\ref{TMD-DOS})
an asymmetric function of the energy.
When a small spin splitting at the conduction band bottom, $\lambda_c$, is taken into consideration
the DOS (\ref{TMD-DOS}) has  $4$ discontinuities: $2$ for the negative  and $2$ for the positive
energies. The positions of these discontinuities are not just at the energies $\pm |\Delta_{\tau \sigma}|$ with
$\tau = \sigma = \pm 1$ and $\tau = -\sigma = \pm 1$ due to the energy shift $E_{\tau \sigma}$.
It is demonstrated that inside the smaller gap there is a region with zero density of states where the dependence of
the entropy per particle on the chemical potential exhibits a huge dip-and-peak structure. The edge of the larger gap corresponds to the discontinuity of the density of states that results in the peak in the dependence
of s on the chemical potential. The large energy gaps in the transition metal dichalcogenides help sustaining the found  resonant features
at rather high temperatures up to $ {100}{K}$.

Since the Seebeck coefficient is related to the temperature derivative of the chemical potential, the strong peaks in the
entropy per particle also indicate the same kind of singularities in the Seebeck coefficient in these materials. This effect is expected at the values of the electronic chemical potential close to the edges of the gaps. The effect has the origin similar to one of the electronic topological transition
\cite{Varlamov1989AP,Blanter1994PR,Sharapov2012PRB}.

\section{Entropy measurements as a tool for detection of topological transitions}\label{germanene}

In this Section based on~\cite{Grassano2018}, we specifically consider germanene. We propose an experimental method for the detection of an electric field induced transition between topological and trivial insulator phases of this  material.
Germanene is a two-dimensional crystal with a buckled honeycomb structure that can be considered as a germanium-based analogue of graphene \cite{Acun2015JPCM}, while it possesses a rather large spin-orbit induced gap in the quasiparticle spectrum.
According to the recent theoretical works, germanene appears to be a natural topological $Z_2$ insulator  \cite{Liu2011PRL,Liu2011PRB}.
Yet it can be brought to the conventional (trivial) insulator phase by applying the external electric field~\cite{Drummond.2012,ezawa2012topological,matthes2014influence,ezawa2015monolayer} perpendicular to its plane which induces a second energy gap owing to the breaking of the inversion symmetry due to the combined effect of the applied field and buckling.
\begin{figure}[!ht]
\centering
\includegraphics[width=10cm]{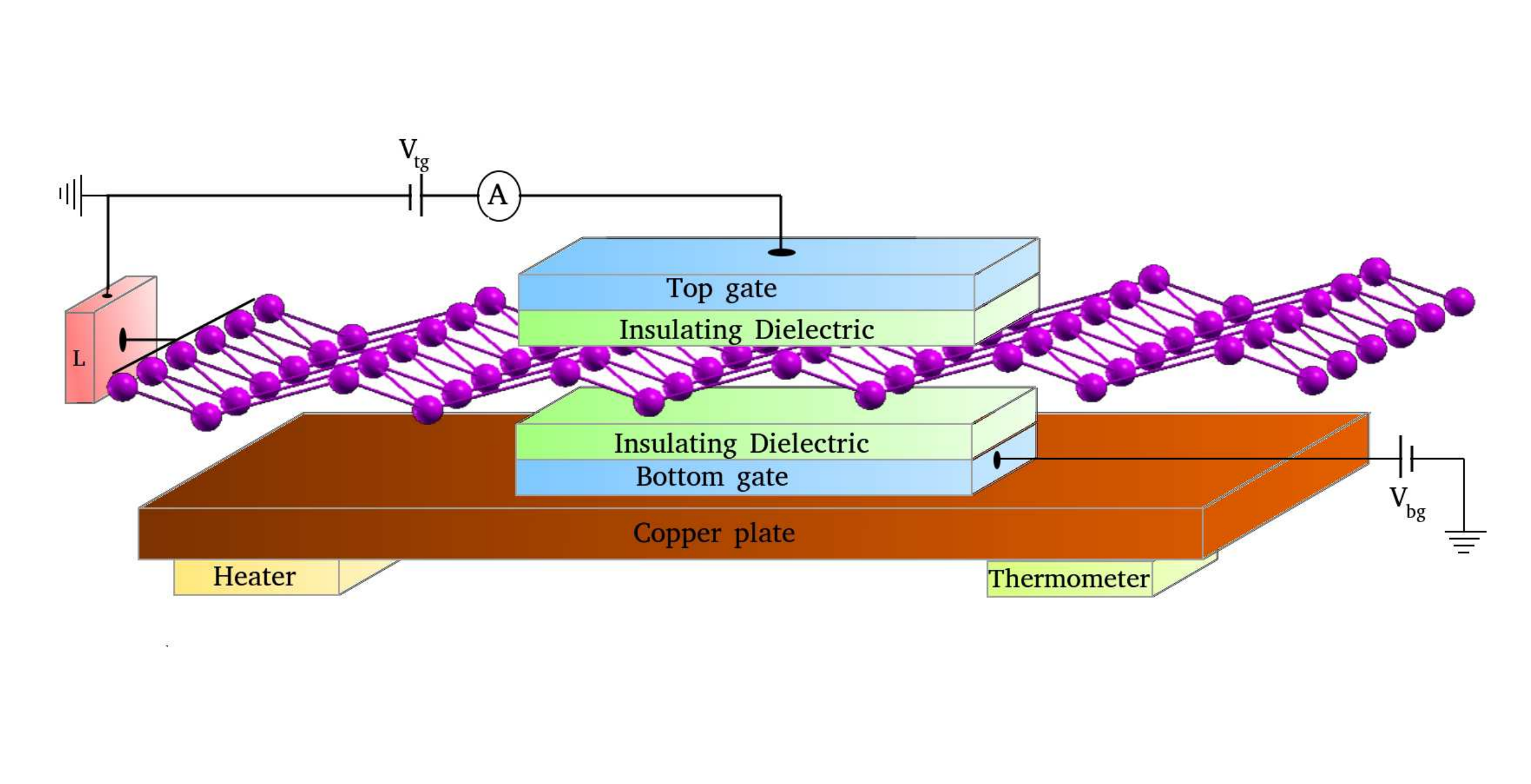}
\caption{(Color online) The schematic of a possible experimental setup to measure entropy per particle, or
$\partial \mu /\partial T$,   in  germanene, or another 2D crystal.
The applied top-gate, $V_{\mathrm{tg}}$,  and bottom-gate, $V_{\mathrm{bg}}$, voltages
allow one to control independently both the chemical potential and the perpendicular
electric field. The sample and the copper sample holder are kept in a thermal contact with a
wire heater, which modulates their temperature and changes  the chemical potential of the sample.
The value $\partial \mu /\partial T$ is directly determined from
the measured recharging current between the crystal and the top electrode \cite{ Pudalov}.
}\label{fig:setup}
\end{figure}

In the present work, we study the entropy per electron dependence on the chemical potential of the electron gas in
a  germanene crystal subjected to the external electric field applied perpendicularly to the crystal plane.
Figure~\ref{fig:setup}  shows a possible experimental setup to measure the entropy per particle, or
$\partial \mu/\partial T$, using dual-gated geometry for band gap engineering (see for example, \cite{Wang2011PRL}).
The top-gate, $V_{\mathrm{tg}}$,  and bottom-gate, $V_{\mathrm{bg}}$, voltages,  are
applied to change the density of carriers and the perpendicular electric field, independently.
Time modulation of the sample temperature changes the chemical potential and
leads to the current flow between germanene sheet and the top gate. The entropy per particle as a function of the chemical potential would be extracted from the recharging current measurements as described in \cite{ Pudalov}.
 
In this Section,  we present the theoretical approach used to model the entropy per particle and explain how we calculate the electronic properties of germanene from the first principles.  
Then we present the results of modeling and show how the entropy per particle dependence on the chemical potential is changing at the transition point between the topological insulator and the trivial insulator phases. Finally, the conclusions are given.

\paragraph{{\it Ab initio} calculations of the DOS. -- }

The {\it ab initio} calculations of the germanene electronic band structure allow extracting the value of topological invariant $Z_2$ as well as the value of the critical electric field $E_c$ where the transition between the topological and trivial phases takes place. 
The same {\it ab initio} calculations provide us by the detailed structure of the one-electron DOS per spin which, in its turn,  yields the required dependence of entropy per particle as the function of the chemical potential.

 Our calculations of DOS and the electronic band structure of germanene are based on the Density Functional Theory (DFT) as implemented in the Quantum Espresso package \cite{espresso.2009,espresso.2017}.
The single-particle Schroedinger equation as formulated by Kohn-Sham~\cite{kohn.sham.1965}:
\begin{equation}
\label{eqn:Kohn-Sham}
\begin{split}
\left(\! -\!\frac{\hbar^2}{2m}\nabla^2\! +\! v_{\mathrm{ext}}({\bf r})\! +\! \int \! \frac{n({\bf r'})}{\left| {\bf r} \!- \!{\bf r'} \right|} d{\bf r'} \!+\! v_{\mathrm{xc}}({\bf r})\!\right)\psi^{\mathrm{K\!S}}_{i,{\bf k}}\!=\!\varepsilon^{\mathrm{K\!S}}_{i,{\bf k}} \psi^{\mathrm{K\!S}}_{i,{\bf k}}
\end{split}
\end{equation}
($v_{\mathrm{ext}}$ is the electron-ions potential and $v_{\mathrm{xc}}$ is the exchange-correlation potential) is solved self-consistently through
the wavefunctions expansion on plane-wave basis sets with use of the periodic boundary conditions.
  For the germanium atoms we use the norm-conserving scheme~\cite{hamann.schluter.1979}, the valence electronic configuration $3d^{10} 4s^2 4p^2$ and the generalized gradient approximation Perdew-Burke-Ernzerhof (GGA-PBE)~\cite{PBE.1996} for the exchange and correlation potential.
  After accurate convergence tests on the total energy results,  an energy cut-off of $90$~Ry has been selected.

We model our 2D crystal  as an infinite $xy$ plane of germanium atoms in the honeycomb geometry.  The theoretical lattice constant for the hexagonal cell, obtained by minimization of the total energy, was found to be $a=4.04 \, \angstrom$ in a low buckling configuration ($\delta_{LB} = 0.68 \, \angstrom$), in agreement with previous results~\cite{prl.ciraci.2009}. Since the periodic boundary conditions are being enforced over all axes, the use of  supercells large enough  to avoid spurious interactions between periodic images is required. After tests over the computed potential and energies, we use a supercell containing $32 \, \angstrom$ of vacuum along the $z$ direction.

 In the absence of the spin-orbit interaction (SO) the germanene spectrum represents a perfect Dirac cone characterized by gapless fermions  with the Fermi velocity of about $0.5 \times 10^6$~m/s.
 By switching on the SO interaction a small gap $\Delta_{SO}$ opens at the K, K$^\prime$ points of the Brillouin zone (BZ), and the band linearity is lost, leading to the appearance of gapped fermions~\cite{SO,Corso.Conte.2005,Conte.Fabris.2008}.
 We obtain a value of the gap of $24$~meV, in agreement with the previous GGA-PBE results. It is slightly below the value of $33$~meV found with use of non-local hybrid exchange and correlation functionals~\cite{matthes2013massive}.

 The gaps in the electronic spectrum can be further modified by applying an external field (bias) perpendicular to the germanene plane.
 This is accounted for by superimposing a sawtooth potential along the z direction of the crystal.
 The properties of the system have been studied for different values of the applied bias, ranging from $0$ to $0.4$~V/\angstrom.
For the DOS calculations a very high energy resolution is needed in order to observe the small differences in low energy features induced by the different electric fields.
For this reason, we used  a refined mesh of $12000 \times 12000 \times 1$  Monkhorst-Pack \cite{Monkhorst.Pack.1976} k-points in the BZ cropped around $\mathbf{K}$($\mathbf{K}^\prime$) with a radius of $0.02 \times 2\pi/a$.

A topological phase transition should be observed at the specific value of the applied field $E=E_c$, where the fundamental electronic gap closes.

As it was mentioned before, at the electric fields below this value, germanene is a topological insulator while above the critical field it becomes a trivial bulk insulator \cite{ezawa2012topological,matthes2014influence}.
In order to prove it, we calculate the topological invariant $Z_2$~\cite{fu.kane.2006}. 
This invariant characterizes phases with nontrivial topological order in time-reversal invariant systems
with a gap and takes only two values: $Z_2=0\,\mbox{mod}\,2$ for a trivial phase, and $Z_2=1\,\mbox{mod}\,2$  for a topologically nontrivial phase. The $Z_2$ invariant divides time-reversal invariant band insulators 
into two classes: ordinary ($Z_2$-even) insulators that can be adiabatically connected to the vacuum  without 
closing a bulk gap, and topological ($Z_2$-odd) ones that cannot be so connected and contain an odd number of the Kramers pairs of counter-propagating edge states leading to the spin Hall effect in two dimensions. 
The $Z_2$-even and $Z_2$-odd phases are separated by a topological phase transition, and the bulk gap  vanishes  at the transition point~\cite{Roy2007}.

For systems with additional inversion symmetry, the invariant $Z_2$ can be calculated using the parity eigenvalues of occupied band states at time-reversal-invariant momenta (TRIM) points (four in two dimensions) in the Brillouin zone~\cite{fu2007topological}.
When the inversion symmetry is broken there are more general approaches for calculating $Z_2$, for example, counting the zeroes of a certain Pfaffian function related to the ground-state wave function \cite{Kane.Mele.2005,fu.kane.2006}, or using the homotopy of the ground-state wave functions in momentum space~\cite{Roy2007}.
Since an electric field perpendicular to the germanene plane breaks the inversion symmetry,  we computed the topological invariant $Z_2$ by applying the method based on the Wilson loop which follows the evolution of the charge centers of  Wannier functions in a plane containing the TRIM points \cite{Soluyanov.Vanderbilt.2011,wison.loop.2011}.  
The numerical implementation of this technique led to the Z2Pack software package~\cite{z2pack.2017} which we used for numerical computation.

\paragraph{Effect of spin-orbit and electric field on the electronic properties. -- }
In the absence of external electric fields, germanene crystals possess the inversion symmetry which, together with time reversal symmetry, leads to the spin degenerate energy bands.
The application of bias brakes the inversion symmetry causing the lifting of the spin degeneracy.
The eigenvalues for each band are computed by solving the Kohn-Sham equation~\eqref{eqn:Kohn-Sham}. The corresponding band structure at K and K$^\prime$ points can be approximated by the relation~\cite{Kane.Mele.2005,Drummond.2012}
\begin{equation}
\label{eqn:band_field}
E_{\eta,\sigma}^{c/v} \ = \ \pm\frac{1}{2}(\Delta_{SO} \ - \sigma\eta\Delta_{el}).
\end{equation}
Here $\Delta_{SO}$ is the spin-orbit splitting of 
$24$~meV, $\Delta_{el}$ is the splitting induced by the electric field, $\eta = \pm 1$ and $\sigma = \pm 1$ are the valley (K and K$^\prime$) and spin ($\uparrow, \downarrow$) subscripts, respectively,  $c$ and $v$ denote the conduction and valence bands corresponding to the $\pm$ signs in the equation.

 As it is seen from Eq.~\eqref{eqn:band_field} four linear bands show up now at K, and four at K$^\prime$ points. They are separated in energy as described by a spin-dependent gap
\begin{equation}
\label{eqn:spin_gap}
\Delta _{\eta,\sigma }= E_{\eta,\sigma}^{c} - E_{\eta,\sigma}^{v}.
\end{equation}
Since the time reversal symmetry is still present,  the relation $E_{\eta,\sigma} = E_{-\eta,-\sigma}$ for the band dispersion holds.
\begin{figure}[!ht]
\centering
  \includegraphics[width=10cm]{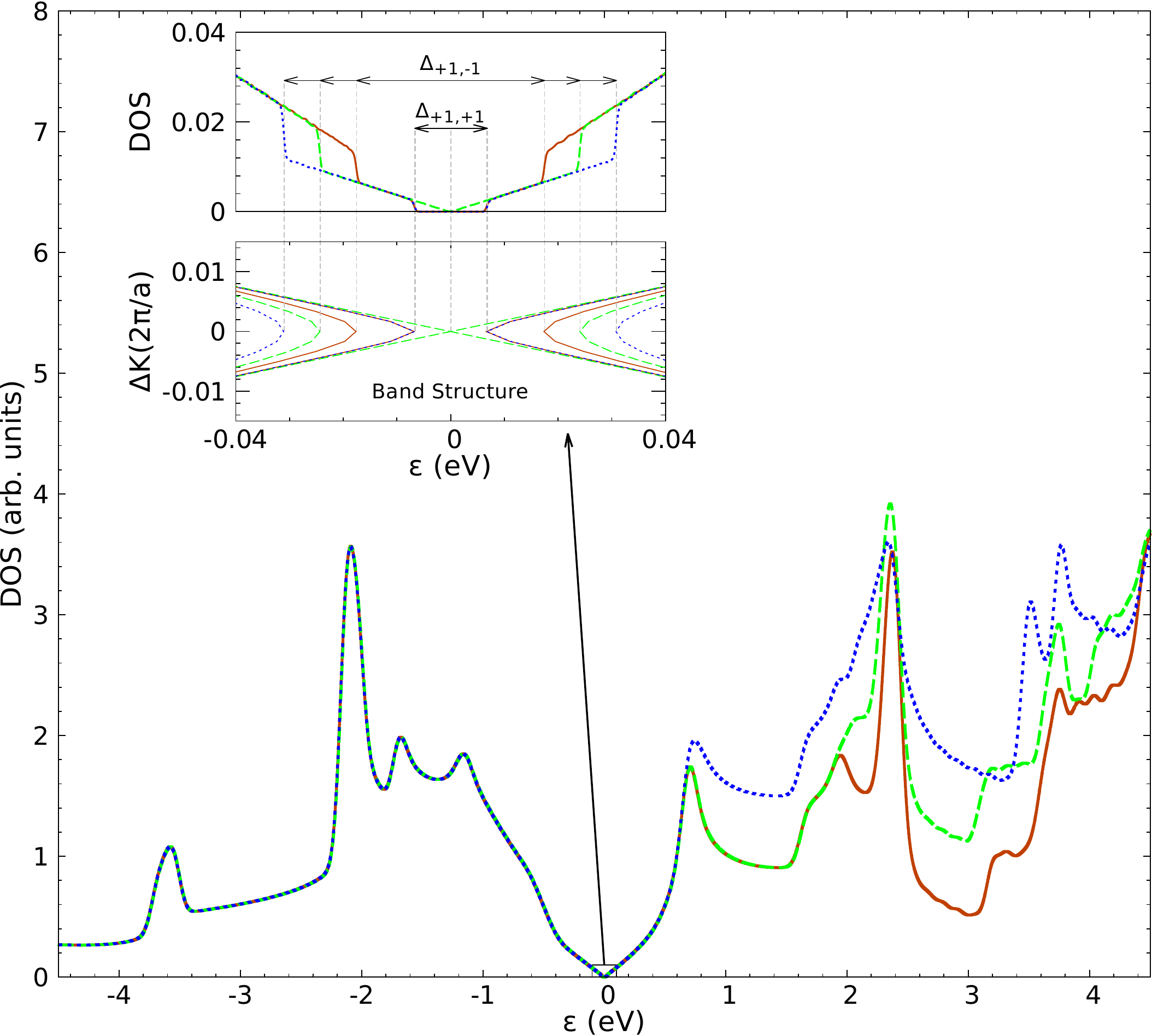}
  \caption{ Density of states computed within the DFT for the external electric fields below/at/above the critical value $E_c$:  $E = 0.10 \,V/\angstrom$ (orange solid line), $0.23 \,V/\angstrom$ (green dashed line) and $0.36 \,V/\angstrom$ (blue dotted line). Zero denotes the Fermi energy.
  Insets: zoom of the DOS near the Fermi energy (upper panel) and the electronic band structure in a close proximity of the high symmetry point K (lower panel).
}\label{fig:DOS+zoom}
\end{figure}

The DOS is   calculated using Eq.~(\ref{eqn:DOS}). The corresponding results are shown in Fig.~\ref{fig:DOS+zoom} for a wide energy range. A broadening of the $\delta$-function in Eq.~\eqref{eqn:DOS} of $50$~meV ($0.3$~meV for the upper inset) has been used.
We can observe that the low energy DOS exhibits the expected linear behavior for the unbiased germanene. The applied electric field leads to the appearance of the gap-like feature in the DOS.
\begin{figure}[!ht]
  \includegraphics[width=\linewidth]{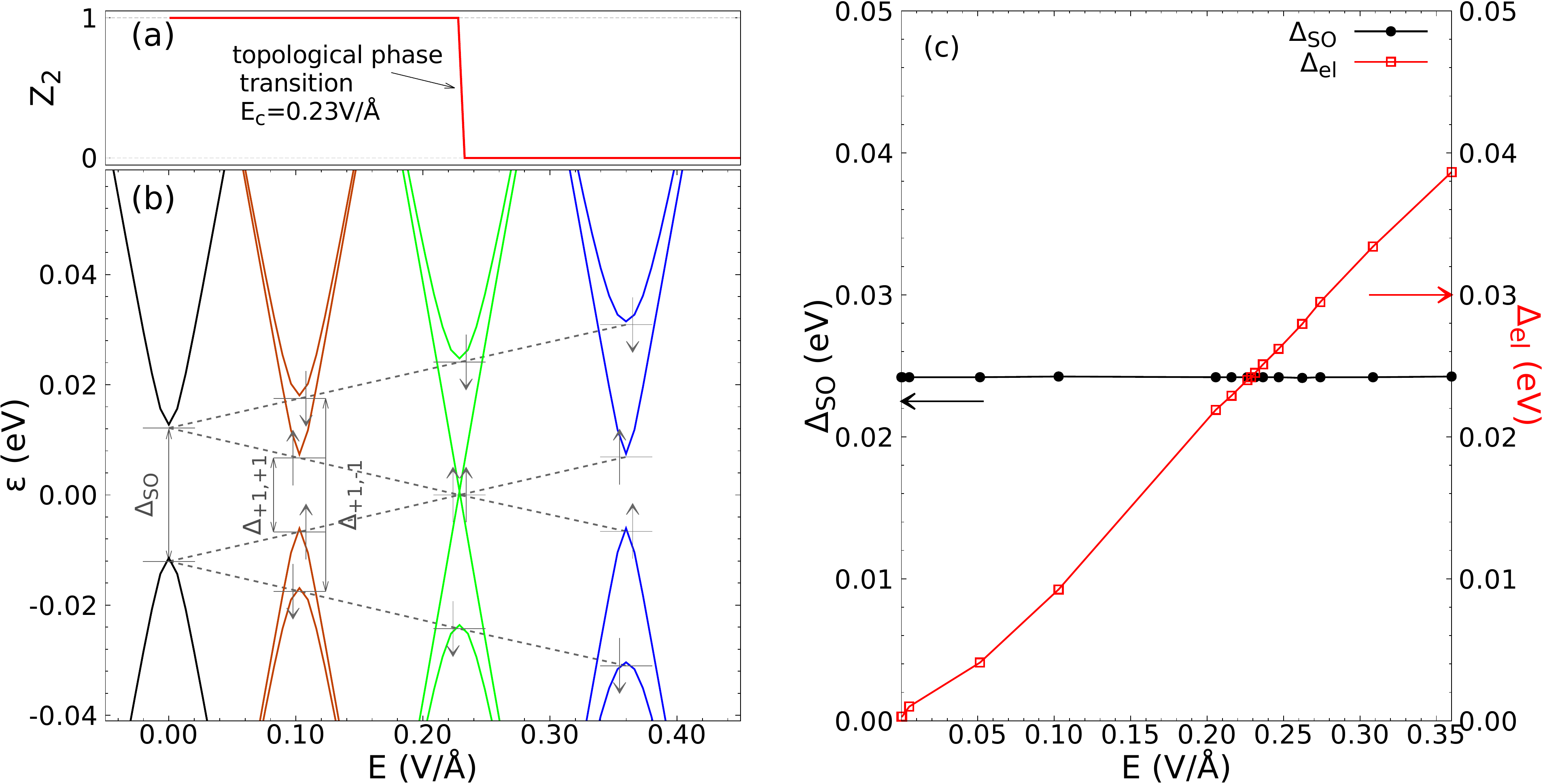}
  \caption{
  a) Computed topological invariant ($Z_2$) at different values of the external electric field.
  b) The electronic band structure near  K-point of the Brillouin zone computed at different values of the external electric field.
  c) The variation of the spin-orbit ($\Delta_{SO}$) and electric field($\Delta _{el}$) induced splittings with the increase of the external electric field.}\label{fig:Gaps2}
\end{figure}

By using Eq. \eqref{eqn:band_field} one can obtain the values of $\Delta_{SO}$ and $\Delta_{el}$ for different electric fields as shown in Fig.~\ref{fig:Gaps2}c.
It can be seen that the former is independent of the field $E$ and remains equal to $24$~meV, while the latter shows a linear dependence on $E$. The splitting $\Delta_{el}$ reaches the value of  $\Delta_{SO}$ at 
$E_c=0.23$~V/\angstrom,  which marks the critical field for the topological phase transition.

We can see in Fig.~\ref{fig:Gaps2}~(c) that the electronic gap $\Delta_{el}$ depend linearly on the applied electric field.
For the values of the field below the critical one ($E < E_c$) the smallest (fundamental) gap decreases, until it closes up completely at $E = E_c$.
In this regime, germanene is a topological insulator, while, for larger field values ($E > E_c$), the smallest  electronic gap opens up again,
and germanene  becomes a trivial insulator.
In what concerns the topological state of germanene we found that $Z_2=1$  for all values of electric field below $E_c$ and $Z_2=0$ otherwise, as
shown in Fig.~\ref{fig:Gaps2}~(a).
%

\paragraph{Dependence of the entropy per particle on the applied electric field. --}

As it has been shown in the previous Sections, see also~\cite{Tsaran2017}, in the case of a crystal characterized by two nonzero energy gaps,
the dependence $s(\mu)$ exhibits two distinct structures in both the electron, $\mu >0$, and hole, $\mu <0$, doped regions.
The first one is a giant resonant feature in the vicinity of zero chemical potential, and the second one is a spike of the height
$s =  2 \ln 2/3$ at the edge of the larger gap. These resonances are apparent in the low temperature limit.

\begin{figure}[!ht]
\centering
\includegraphics[width=10cm]{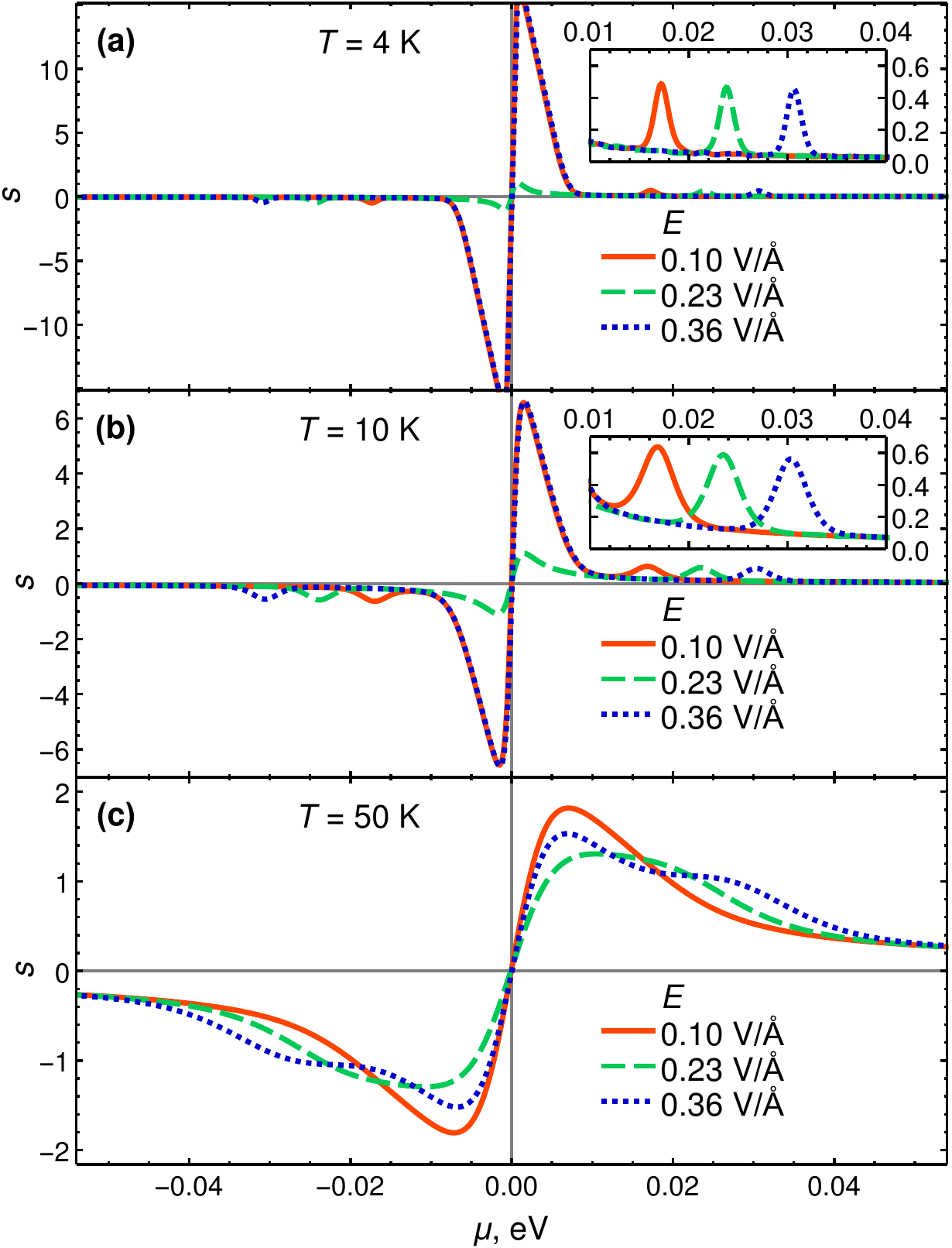}
\caption{(Color online) The entropy per electron $s$ vs the chemical potential $\mu$ in eV in the vicinity of the Dirac point for three values of the electric field $E$:
(a) $T=4$~K, (b) $T = 10$~K,  (c) $T = 50$~K.
Insets in (a) and (b) show the zoomed domains with the entropy spikes of the height $s = 2 \ln2/3$ at low temperatures.}
\label{fig:Spikes}
\end{figure}

Our results for the entropy per particle $s$ for germanene are presented in Fig.~\ref{fig:Spikes}.
The entropy spikes occur both above and below the transition between the topological and
trivial insulator phases, as well as exactly at the transition point, $E_{c} = 0.23$~V/\AA~(see the insets of Figs.~\ref{fig:Spikes}~(a), (b)).
The most prominent result of the present Section is that the strong resonant feature of the entropy per particle in the close vicinity
of the Dirac point, $\mu =0$, is nearly fully suppressed at the transition point, $E_c$,  while it occurs for values of the
electric field below and above it.

It is important to note that the appearance of the second step in the electronic density of states due to the lifted spin degeneracy of the  Brillouin zone has a dramatic effect on $s$.
In particular, the resonant feature in the vicinity of zero chemical potential is strongly pronounced
if the DOS exhibits two steps (see upper inset in Fig.~\ref{fig:DOS+zoom}). Indeed, for
$|\mu| \ll T \ll \Delta_{+1,+1}$  it was obtained~\cite{Tsaran2017} (see Eq.~(\ref{mu=0})) that
\begin{equation}
s(T, \mu ,\Delta_{+1,+1} ) \simeq \frac{\mu \Delta_{+1,+1}}{2T^2} .
\end{equation}
For the critical field $E=E_c$ the gap $\Delta_{+1,+1} = 0$, so that the DOS exhibits
only one step and in accordance with Eq.~(\ref{s(T,0,mu)}) 
\begin{equation}
s(T,\mu, 0 ) \simeq \frac{\mu}{T}, \quad   |\mu| \ll T.
\end{equation}
Clearly, at very low temperature the peak at finite $\Delta_{+1,+1}$ is much stronger than
one for  $\Delta_{+1,+1}=0$.

The disappearance of the characteristic entropy resonance can be considered as a signature of the topological phase transition in germanene.
We are confident that this analysis would help extracting the important band parameters of topological 2D crystals from the recharging current measurements.

\begin{figure}[!ht]
\includegraphics[width=\linewidth]{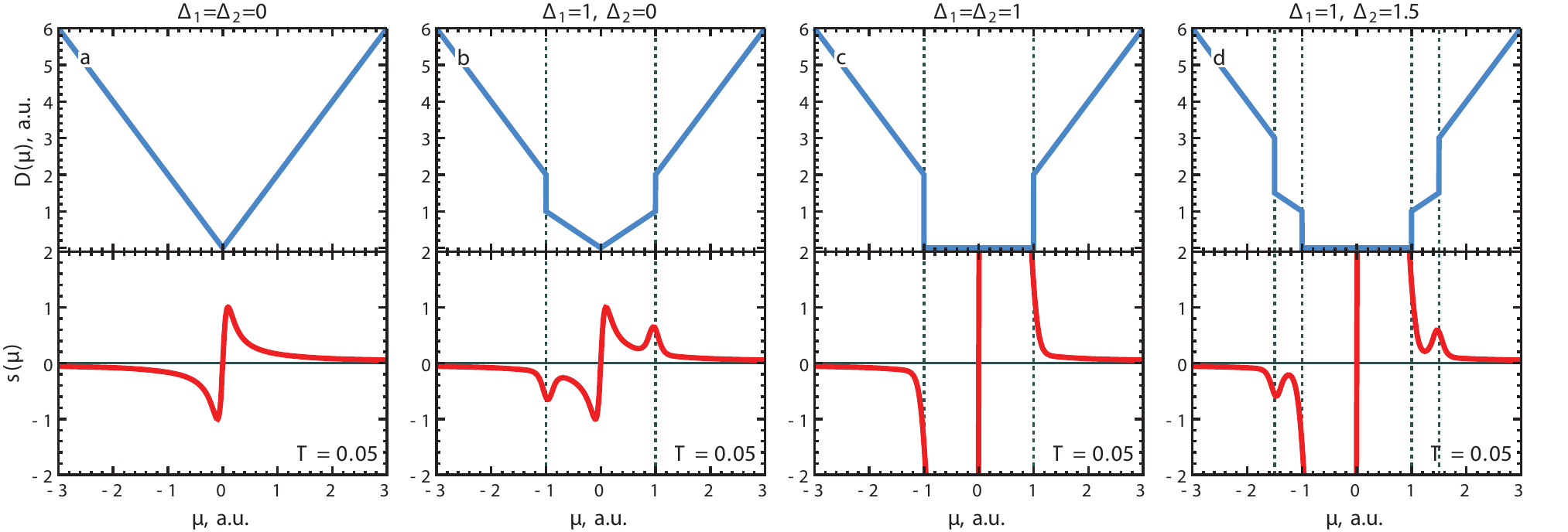}
\caption{(Color online) The correlation between the shape of the DOS shown in the top and behavior of $s(\mu)$ 
shown underneath.  (a) Massless Dirac fermions in graphene. (b) Silicene and others at the point of topological transition. (c)
Two gaps are equal to each other. (d) Two different gaps.
}
\label{fig:explanation}
\end{figure}

The correlation between the shape of the DOS and the behavior of $s(\mu)$ is illustrated in Fig.~\ref{fig:explanation}~(a) is the case of graphene with massless Dirac fermions. Panels (b), (c) and (d) represent  silicene, germanene and related materials with two energy gaps taking different values. Specifically, the panel (b) is for one finite gap and second zero gap, as occurs at the point of the topological transition.  The resonant feature in the vicinity of zero chemical potential is strongly suppressed as compared to the next two panels; panel 
(c) corresponds to the degenerate case where the two gaps are equal, 
as occurs in silicene and germanene for $E=0$ and in the gapped graphene; panel(d) corresponds to two finite gaps.

Finally, we note that it was suggested
that the plasmon modes and Friedel oscillation can be used to detect the topological phase transition 
in silicene and germanene even when the Fermi level does not lie in the band gap~\cite{Chang2014}. 

\paragraph{Van Hove singularities in the 
$s(\mu)$ dependence
on a large energy scale. --}

\begin{figure}[!ht]
\centering
\includegraphics[width=12cm]{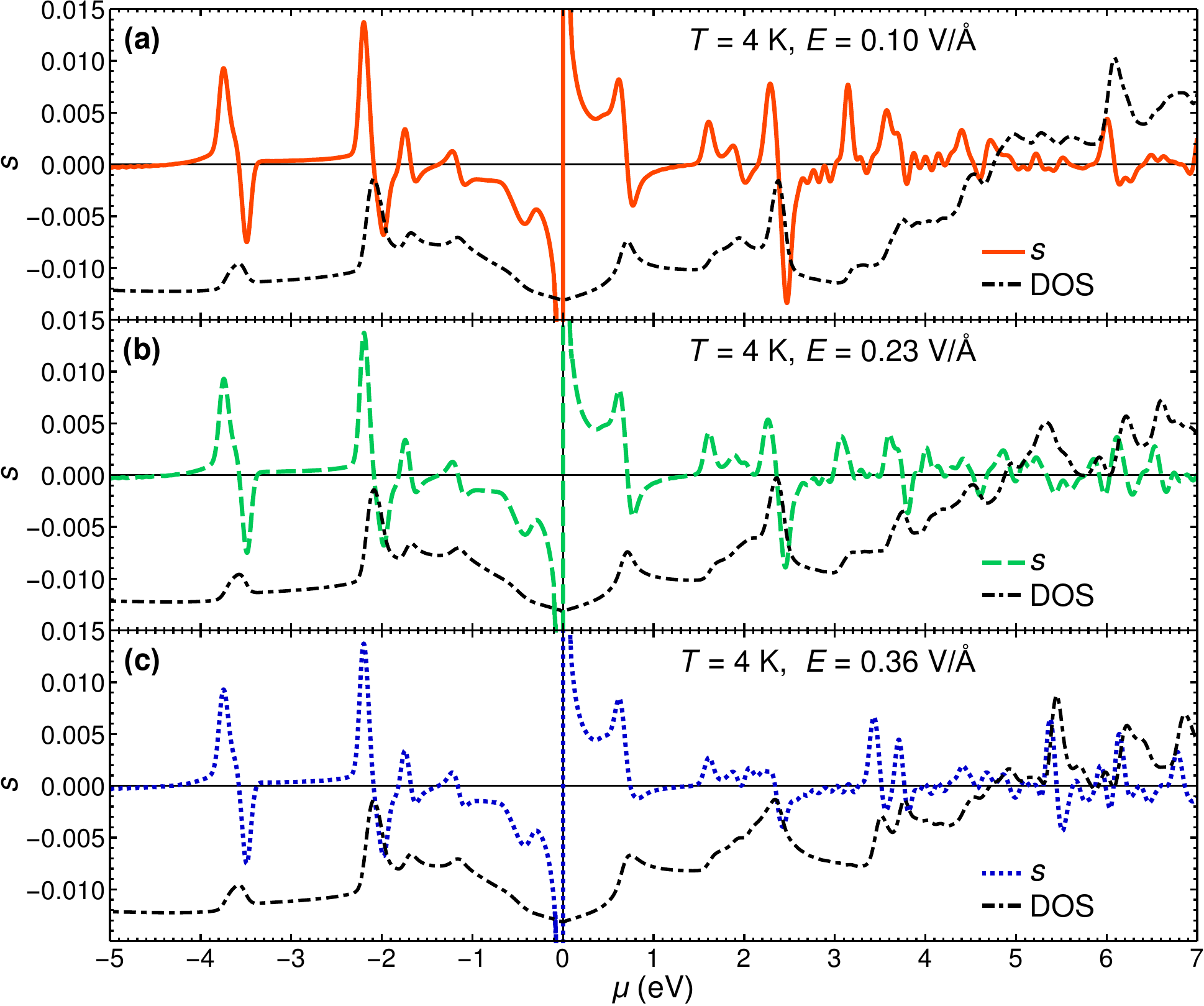}
\caption{(Color online) The entropy per electron $s$ vs the chemical potential $\mu$ in eV for three values of
electric field $E$ for $T= 4 \mbox{K}$:
(a) $E= 0.10\, \mbox{V/\AA}$ (b) $E= 0.23\, \mbox{V/\AA}$ (c) $E= 0.36~\mbox{V/\AA}$. Dashed lines show the
corresponding DOS obtained by the ab-initio calculation. }\label{fig:Spikes-fullrange}
\end{figure}

Scanning of the chemical potential on a large energy scale is challenging from the experimental point of view.
Possibly, this could be done by combining electrostatic and chemical doping.
Nevertheless, the behavior of the entropy per particle at large values of the chemical potential is worth analyzing theoretically as it offers some non-trivial features. Figure \ref{fig:Spikes-fullrange} shows $s(\mu)$ in comparison with DOS for three values of the applied field.
We note that both energy and entropy scales in Fig.~\ref{fig:Spikes-fullrange} are orders of magnitude different from those
in  Fig.~\ref{fig:Spikes}, so that the high-energy features in $s(\mu)$ are several orders of magnitude weaker then the resonant feature at zero chemical potential discussed above. In particular,
the disappearance  of the resonant feature at the critical field $E_c$  cannot be recognized in Fig.~\ref{fig:Spikes-fullrange}~(b)
due to the small range of shown values of $s$.
Nevertheless, these features represent a significant interest. Using
Eq.~(\ref{entropy-part-b}) it is easy to see that the extrema of the dependence of $D(\mu)$ are converted to the zeros of $s(\mu)$.
Indeed, assuming that $D(\mu)$ is a smooth function one may expand it near the extremum of $\mu_{\mathrm{ex}}$ and
obtain that $s(\mu) \varpropto D^\prime(\mu_{\mathrm{ex}})$ \cite{Tsaran2017}, where the derivative
$D^\prime(\mu_{\mathrm{ex}})$ changes sign at the extrema of $\mu_{\mathrm{ex}}$.
Since van Hove singularities correspond to the sharp peaks in the DOS, they show up in the dependence
$s(\mu)$ as the strong positive peak  and negative dip structures.
Note that the discussed above giant negative dip and positive peak structure near zero chemical potential may be interpreted as a signature for the negative $V$ and $U$-like shape peaks of the DOS.
This demonstrates that the experimental investigation of $s(\mu)$ in a wider range of energies can be useful for
tracing out the van Hove singularities of DOS and their evolution in the perpendicularly applied electric field.

To conclude, in this Section, we have studied fingerprints of the topological phase transitions
and DOS singularities in two-dimensional materials with use of the ab-initio calculations. We show that the entropy per particle dependence on the chemical potential is highly sensitive to DOS. In particular, at the critical field corresponding to the transition point between topological and trivial insulator phases, the strong resonant feature of the entropy per particle at zero chemical potential disappears. Moreover, at the Van Hove singularities of DOS, the entropy per particle passes through zero (dip-peak features). Based on these theoretical findings, we propose an experimental method of detection of the critical transition points and density of states singularities in novel structures and materials.

\section*{Conclusions}
In this review paper we have considered several examples of two-dimensional electronic systems undergoing topological transitions of various types. In particular, we discussed the quantum confined 2DEG in semiconductor quantum wells, the gas of Dirac electrons in the gapped graphene, electrons and holes in silicene, germanene and transition metal dichalcogenides. For each of this systems we analyzed the dependence of the entropy per particle on the electronic chemical potential and thoroughly discussed the peculiarities of this function in the vicinity of topological transition points. We have shown, specifically, that: (i) the entropy per particle exhibits quantized peaks at the intersection points of the chemical potential and size quantization levels in a 2DEG with parabolic energy subbands. The amplitude of each peak is set by the size quantization number of the corresponding subband and the combination of fundamental constants; (ii) in gapped Dirac materials characterized by two energy gaps, the entropy per particle shows the analogous to 2DEG spikes at the resonances of the electron (hole) chemical potential and the gap edges; 
(iii) monoatomic layers of transition metal dichalcogenides demonstrate similar to low-buckled Dirac materials entropy features, however the symmetry between electron and hole parts of the entropy dependence on the chemical potential is broken in TMDCs;
(iv) A very strong Lorentz-shape feature of the entropy per particle dependence on the chemical potential is found at zero chemical potential (corresponding to the middle of the energy gap) for any of considered system; (v) we show that this zero-energy resonance is abruptly suppressed at the point of transition from the topological to the trivial insulator phase in germanene.  This observation can constitute a smoking gun for such kind of topological phase transition. 

We emphasize that the theory predictions (i-iv) still need their experimental verification. An appropriate experimental technique for the measurement of the entropy per particle in 2D crystals would be the measurement of recharging currents in the planar capacitor geometry, as described in Ref.~\cite{Pudalov}. The formalism presented here can be naturally extended to the studies of 2D crystals subjected to external magnetic fields. Some useful expressions for the temperature derivative of the chemical potential in graphene and intercalated graphite subjected to the magnetic field can be found in Ref.~\cite{Lukyanchuk2011}.

The effects of electron-electron interaction on the effective DOS and,
consequently, on the entropy per particle remain out of the scope of this review paper,
while we realize that in specific two-dimensional crystal systems these interactions may
play an important role. All the results presented here are obtained with use of a single-electron
DOS, that may be considered as the first order approximation.
Next order corrections may be taken into account in many ways, which would constitute
a separate significant piece of study. We underline, however, that we did consider here the 	homogeneous broadening effect on DOS (such as the thermal broadening). We believe that the 	temperature may be considered as a fitting parameter in many cases, as the real temperature of 	the electron gas may be quite different from the crystal lattice temperature. Varying this effective 	temperature one should be able to account also for other broadening factors, to some extent.

Finally, we note that the entropy per particle is an important characteristic of any many-body system, that is not yet sufficiently well understood and experimentally studied. The goal of this review is to attract attention of the scientific community to the surprising behavior of the entropy per particle in the vicinity of topological transitions in various two-dimensional system and to stimulate its further experimental and theoretical studies.

\section*{Acknowledgements}

We thank A.O. Slobodeniuk for illuminating discussion. We acknowledge the support from the HORIZON 2020 RISE "CoExAN" project (GA644076).
A.V.K. acknowledges support from the St-Petersburg State University for the research grant 11.34.2.2012.
S.G.Sh. and V.P.G. acknowledge a partial support by the National Academy of Sciences of Ukraine (projects No. 0117U000236 and 0116U003191) and by its Program of Fundamental Research of the Department of Physics and Astronomy (project No. 0117U000240).

\begin{appendices}
\numberwithin{equation}{section}
\section{Derivation of Equations~(\ref{dndT}) and (\ref{dndm})}\label{app_a}

Using Eqs.~(\ref{ngen}) and (\ref{theta}) we can cast the derivatives $(\partial n/ \partial \mu)_T$  and $(\partial n/ \partial T)_\mu$ in the form:
\begin{eqnarray}
\left(\frac{\partial n}{\partial\mu}\right)_T&=& \frac{m*}{2\pi\hbar^2}\sum_j\int\limits_{-\infty}^\infty
\frac{dz}{\cosh^2z}\left[\frac{1}{2}+\frac{1}{\pi}\arctan(a z+b_j)\right], \\
\left(\frac{\partial n}{\partial T}\right)_\mu&=& \frac{m*}{\pi\hbar^2}\sum_j\int\limits_{-\infty}^\infty
\frac{z \, dz}{\cosh^2z}\left[\frac{1}{2}+\frac{1}{\pi}\arctan(a z+b_j)\right],
\end{eqnarray}
where $a \equiv 2T/\gamma$, $b_j\equiv (\mu-E_j)/\gamma$.
Therefore, one is left with the integrals
\begin{equation}
I(a,b)=\int\limits_{-\infty}^\infty\frac{dz\,z}{\cosh^2z}\arctan(az+b),\quad  
J(a,b)=\int\limits_{-\infty}^\infty\frac{dz}{\cosh^2z}\arctan(az+b)
\end{equation}
where  $a > 0$. 

As an example, let us consider the integral $I(a,b)$, which is even function in $b$.
Clearly,
\begin{equation} \label{cond_a1}
I(a,b \to \infty)=0.
\end{equation}
It is worth evaluating the derivative
\begin{equation}
\frac{\partial I(a,b)}{\partial b}=\int\limits_{-\infty}^\infty\frac{dx}{\cosh^2x}\frac{x}{(ax+b)^2+1}.
\end{equation}
Since
\begin{equation}
\tanh(x)=2\sum\limits_{n=0}^\infty\frac{x}{c_n^2+x^2},\quad c_n=\pi(n+1/2),
\end{equation}
by differentiation over $x$ one obtains
\begin{equation}
\frac{1}{\cosh^2x}=2\sum\limits_{n=0}^\infty\left[\frac{2c_n^2}{(c_n^2+x^2)^2}-\frac{1}{c_n^2+x^2}\right], \quad c_n=\pi(n+1/2).
\label{series}
\end{equation}
Therefore,
\begin{eqnarray}
\frac{\partial I(a,b)}{\partial b}&=&2\sum\limits_{n=0}^\infty\int\limits_{-\infty}^\infty\frac{dx\,x}{(ax+b)^2+1}
\left[\frac{2c_n^2}{(c_n^2+x^2)^2}-\frac{1}{c_n^2+x^2}\right]\nonumber\\
&=&2\pi b\sum\limits_{n=0}^\infty\frac{1+b^2-a^2c_n^2}{[b^2+(1+ac_n)^2]^2}
=\frac{2}{\pi a^2}{\rm Im}\left[(1-ib)\Psi'\left(\frac{1}{2}+\frac{1-ib}{\pi a}\right)\right].
\label{derivative_I_over_b}
\end{eqnarray}
Integrating Eq.~(\ref{derivative_I_over_b}) over $b$ we get
\begin{equation}
I(a,b)=\frac{2}{a}{\rm Re}\left[(1+ib)\Psi\left(\frac{1}{2}+\frac{1+ib}{\pi a}\right)-
\pi a\ln\Gamma\left(\frac{1}{2}+\frac{1+ib}{\pi a}\right)\right]+C(a).
\end{equation}
At $b \gg 1$ we find that
\begin{equation}
I(a,b)\simeq\frac{2}{a}{\rm Re}\left[i b+\left(1-\frac{\pi a}{2}\ln(2\pi)-\frac{i\pi^2a^2}{12b}+O\left(\frac{1}{b^2}\right)\right)\right]
+C(a).
\end{equation}
Using the equality (\ref{cond_a1}) one obtains
\begin{equation}
C(a)=\frac{2}{a}\left(-1+\frac{\pi a}{2}\ln(2\pi)\right).
\end{equation}
The results of the aforementioned procedure and a similar procedure for $J(a,b)$ read as:
\begin{eqnarray}
I(a,b)&=&\frac{2}{a}{\rm Re}\left[(1+ib)\Psi\left(\frac{1}{2}+\frac{1+ib}{\pi a}\right)-1-
\pi a\ln\Gamma\left(\frac{1}{2}+\frac{1+ib}{\pi a}\right)+\frac{\pi a}{2}\ln(2\pi)\right],
\\
J(a,b)&=&2 \,{\rm Im}\left[\Psi\left(\frac{1}{2}+\frac{1+i b}{\pi a}\right)\right].
\end{eqnarray}
Using the above auxiliary integrals one easily obtains Eqs.~(\ref{dndT}) and (\ref{dndm}).
The limiting case $\gamma \to 0$ follows from the properties of the $\Psi$-function:
\begin{equation} \label{lim01}
{\rm Im}\left[\Psi\left(\frac{1}{2}+i x\right)\right]=\frac{\pi}{2}\tanh(\pi x), \quad 
\Psi(z)\simeq \ln z +O(\frac{1}{z}),\quad |z|\gg 1.
\end{equation}

\section{Gapped Dirac materials: Details of calculations} \label{appendixB}
\numberwithin{equation}{section}

 \paragraph{Relationship between the carrier density and carrier imbalance. --}

 In a relativistic theory, for example, in  QED the number of electrons or positrons is not conserved, while
 a conserving number operator is needed to build the statistical density matrix~\cite{Kapusta.book}.
 In QED, the conserved quantity if the difference of the numbers of positively and negatively charged particles: electrons and positrons.

 In the Dirac materials the ``relativistic'' nature of carriers is encoded in the
 symmetric DOS function, $D(\varepsilon) = D(-\varepsilon)$. Accordingly,
 it is convenient to operate with the difference between the densities of electrons and holes instead of the total
 density of electrons~\cite{Gusynin2004PRB,Sharapov2015JPA}. The difference  is given by
 \begin{equation}
 \label{number-rel}
 n(T,\mu)  =  \int_{-\infty}^{\infty } d \varepsilon D(\varepsilon) [ f_{F D}(\varepsilon - \mu) \theta(\varepsilon)
  -[1- f_{F D}(\varepsilon - \mu) ] \theta(-\varepsilon)]
  = -\frac{1}{2} \int_{-\infty}^{\infty } d \varepsilon D(\varepsilon) \tanh \frac{\varepsilon-\mu}{2T}.
 \end{equation}
 The last equation can be rewritten in the form of Eq.~(3).
 One can verify that the carrier imbalance $n(T,\mu)$ and the total carrier density $n_{\mathrm{tot}}(T,\mu) $
 are related by the expression
 $n(T,\mu)  = n_{\mathrm{tot}}(T,\mu) - n_{\mathrm{hf}}$, where $n_{\mathrm{hf}}$ is the density of particles
 for a half-filled band (in the lower Dirac cone)
 $
 n_{\mathrm{hf}} =  \int_{-\infty}^{\infty } d \varepsilon D(\varepsilon) \theta(-\varepsilon).
 $
 Consequently, there is no difference whether the entropy per particle 
 is defined
 via the total carrier density $n_{\mathrm{tot}}$ or the carrier imbalance $n$.

 \paragraph{Expressions for $ \partial n /\partial T$ and $\partial n /\partial \mu$. --}

 The first temperature derivative in Eq.~(\ref{fullderiv}) depends on whether the chemical potential
 $\mu$ hits the discontinuity of the DOS $D(\varepsilon)$ given by Eq.~(\ref{DOS-general}).
 Differentiating Eq.~(\ref{number-general}) over the temperature one obtains
 \begin{equation}
  \frac{\partial n(T,\mu)}{\partial T} =  \frac{{\rm sign}(\mu)}{4T}\int_{-\infty}^\infty d\varepsilon D(\varepsilon)
  \left[
 \frac{\varepsilon-|\mu|}{2T}\frac{1}{\cosh^2\frac{\varepsilon-|\mu|}{2T}}-\frac{\varepsilon+|\mu|}{2T}\frac{1}{\cosh^2
 \frac{\varepsilon+|\mu|}{2T}}\right].
 \end{equation}
 Changing the variable $\varepsilon=2Tx\pm|\mu|$ in two terms and changing the limits of integration, one obtains
 \begin{equation}
  \frac{\partial n(T,\mu)}{\partial T} =
  {\rm sign}(\mu)\int\limits_{0}^\infty dx\left[D(|\mu|+2Tx)-D(|\mu|-2Tx)\right]
 \frac{x}{\cosh^2x}.
 \end{equation}
 If the DOS $D(E)$  has a continuous derivative at the point $E = |\mu|$, where $\Delta_i < |\mu| < \Delta_{i+1}$,
 one can expand $D(|\mu|+2Tx)-D(|\mu|-2Tx)\simeq 4TxD^\prime(|\mu|)$. Then integrating over $x$ we arrive at Eq.~(\ref{dNdT})
 \begin{equation}
 \frac{\partial n(T,\mu)}{\partial T}  \simeq4T{\rm sign}(\mu)D^\prime(|\mu|)\int\limits_{0}^\infty \frac{x^2 \, d x}{\cosh^2x}
 ={\rm sign}(\mu)D^\prime(|\mu|)\frac{\pi^2}{3}T.
 \end{equation}
 On the other hand,  at the discontinuity points $\mu=\pm\Delta_J$ at $T \to 0$, we arrive at Eq.~(\ref{dNdT-disc}).

 The second derivative in Eq.~(\ref{fullderiv}) in the zero temperature limit is just the DOS.
 Indeed, we have
 \begin{equation}
 \label{mu-derivative}
 \frac{\partial n(T,\mu)}{\partial\mu} =\frac{1}{8T}\int_{-\infty}^\infty d\varepsilon D(\varepsilon)\left[
 \frac{1}{\cosh^2\frac{\varepsilon+\mu}{2T}}+\frac{1}{\cosh^2\frac{\varepsilon-\mu}{2T}}\right]
  \to  D(\mu),\qquad T\to 0.
 \end{equation}
 This is because
 $(1/4T) \cosh^{-2} (x/2T) \rightarrow\delta(x)$ for $x\to0$. Substituting the DOS given by Eq.~(\ref{DOS-general}) to
 Eq.~(\ref{mu-derivative}) we arrive at Eq.~(\ref{dNdmu}).

The carrier imbalance for a gapped graphene is given by Eq.~(\ref{number-graphene}). The corresponding derivatives
 are given by Eqs.~(\ref{derivative-T-2nd-a}) and (\ref{derivative-mu-a}).
%
 \begin{eqnarray}
 \label{derivative-mu}
  \left( \frac{\partial n}{\partial \mu }\right) _{T} &=&  \frac{2}{\pi \hbar
 ^{2}v_{F}^{2}}\left[ \frac{\Delta }{2}\left( \tanh \frac{\mu -\Delta }{2T}%
 -\tanh \frac{\mu +\Delta }{2T}\right)
 + T\left( \ln \left( 2\cosh \frac{\mu -\Delta }{2T}\right) +\ln
 \left( 2\cosh \frac{\mu +\Delta }{2T}\right) \right) \right]
 \\
  \left( \frac{\partial n}{\partial T} \right)_\mu &= & \frac{2}{ \pi \hbar^2 v_F^2}
 \left[ 2\Delta \ln \frac{1+\exp \left( \frac{\mu -\Delta }{T}%
 \right) }{1+\exp \left( -\frac{\mu +\Delta }{T}\right) }
  +2T\mbox{Li}_{2} \left( -e^{-\frac{\mu +\Delta }{T}}\right)
 -2T\mbox{Li}_{2}\left( -e^{\frac{\mu -\Delta }{T}}\right) \right. \nonumber\\
  &&\left. -\mu \ln \left( 2\cosh \frac{\mu -\Delta }{2T}\right) -\mu \ln \left( 2\cosh \frac{%
 \mu +\Delta }{2T}\right)
 + \frac{\Delta }{T}\frac{\mu \sinh (\Delta /T)+\Delta \sinh \mu /T}{%
 \cosh \Delta /T+\cosh \mu /T}\right] .
 \label{derivative-T-2nd}
 \end{eqnarray}
 
 Equations~(\ref{s-mu<Delta})--(\ref{2ln2-T}) and (\ref{2ln2-T-2nd}) are obtained using the low-temperature expansions of the derivatives,
 Eqs.~(\ref{derivative-T-2nd-a}) and (\ref{derivative-mu-a}).

We also provide the corresponding expressions for the zero gap graphene and 2DEG.
In  the case $\Delta =0$ Eq.~(\ref{number-graphene}) reduces to
\begin{equation}
 \label{number-graphene-0}
 n(T,\mu) = \frac{2T^2}{\pi \hbar^2 v_F^2}
 \left[ \mbox{Li}_2 \left(-e^{- \frac{\mu}{T}} \right) - \mbox{Li}_2 \left(-e^{ \frac{\mu}{T}} \right) \right].
 \end{equation}
 Using Eq.~(\ref{fullderiv}) we obtain the general expression
 \begin{equation}
 \label{derivative-Delta=0}
 \left( \frac{\partial \mu}{\partial T} \right)_n  = \frac{\mu}{T}-  \frac{1}{\ln \left (2 \cosh \frac{\mu}{2T} \right)}
 \left[ \mbox{Li}_2 \left(-e^{- \frac{\mu}{T}} \right) - \mbox{Li}_2 \left(-e^{ \frac{\mu}{T}}\right) \right].
 \end{equation}

 In the 2DEG in the presence of Zeeman splitting considered in the Supplementary material of \cite{ Pudalov}
 the carrier density reads
 \begin{equation}
 \label{number:eq}
 n(\mu,T)=\frac{m}{4\pi}T\left[\ln\left(1+e^{(\mu+Z)/T}\right)+\ln\left(1+e^{(\mu-Z)/T}\right)\right].
 \end{equation}
 Here $Z$ is the Zeeman splitting energy and $m$ is the carrier mass.
 One can show that the entropy per particle in this case also obeys
 the  quantization rule
 \begin{equation}
 \label{s-minus-Z}
 \left. \frac{\partial S}{\partial n} \right|_{\mu=-Z}=2\ln2,
 \quad \left. \frac{\partial S}{\partial n} \right|_{\mu=Z}=\frac{2\ln2}{3}, \quad T\to 0.
 \end{equation}
 \end{appendices}
 

\end{document}